\documentclass[12pt]{article}
\pdfoutput=1
\usepackage{makeidx}
\makeindex
\usepackage[a4paper]{geometry}
\usepackage{jheppub,amsmath,amssymb,amsfonts,amsxtra,mathrsfs,graphics,graphicx,amsthm,epsfig,ytableau,bm,longtable,float,color,tikz,mathtools,xfrac,footnote,rotating,lscape}
\restylefloat{table}
\usepackage{dsfont}
\usepackage{multicol}
\usepackage{lipsum}
\usepackage{textgreek}
\usetikzlibrary{decorations.pathmorphing}
\usetikzlibrary{decorations.markings}
\usetikzlibrary{quotes,arrows.meta}
\usetikzlibrary{arrows, decorations.markings, calc, fadings, decorations.pathreplacing, patterns, decorations.pathmorphing, positioning}
\usepackage{tikz-cd}

\usepackage{fixmath} 
\usepackage{scalerel}
\newlength\bshft
\def\fakebold#1{\ThisStyle{\ooalign{$\SavedStyle#1$\cr%
  \kern-\bshft$\SavedStyle#1$\cr%
  \kern\bshft$\SavedStyle#1$}}}

\usetikzlibrary{positioning,shapes}
\usetikzlibrary{chains}
\usetikzlibrary{arrows,fit,decorations.pathreplacing}
\tikzstyle{every picture}+=[remember picture]
\tikzstyle{na} = [baseline=-.5ex]

\usepackage{empheq}
\usepackage{multirow}
\usepackage{booktabs}
\usepackage[american]{babel}

\usepackage[latin1]{inputenc}

\usepackage{array,booktabs}
\def\mystrut{\rule{0pt}{1.25\normalbaselineskip}}

\makeatletter
\newcommand{\vast}{\bBigg@{1}}
\newcommand{\Vast}{\bBigg@{5}}
\makeatother

\usepackage{hyperref}

\numberwithin{equation}{section}

\newcommand{\cf}{\textit{cf.}}
\newcommand{\eg}{\textit{e.g.}}

\newcommand{\ie}{\textit{i.e.}}

\newcommand{\ii}{\mathrm{i}}
\newcommand{\?}{\;\!}

\numberwithin{equation}{section}

\newcommand{\be}{\begin{equation}} \newcommand{\ee}{\end{equation}}
\newcommand{\bea}{\begin{equation} \begin{aligned}} \newcommand{\eea}{\end{aligned} \end{equation}}

\newcommand{\Iprod}[2]{\langle {#1}, {#2} \rangle}

\def\U{\mathrm{U}}
\def\SO{\mathrm{SO}}

\def\SU{\mathrm{SU}}

\def\Sp{\mathrm{Sp}}

\def\USp{\mathrm{USp}}

\def\u{\mathsf{u}}
\newcommand{\ex}{{\mathrm{e}}}
\newcommand{\rd}{\mathrm{d}}

\newcommand{\Vol}{\mathrm{Vol}}

\newcommand{\wt}{\widetilde}

\newcommand{\cB}{\mathcal{B}}
\newcommand{\cC}{\mathcal{C}}

\newcommand{\cE}{\mathcal{E}}
\newcommand{\cF}{\mathcal{F}}

\newcommand{\cH}{\mathcal{H}}
\newcommand{\cI}{\mathcal{I}}
\newcommand{\cJ}{\mathcal{J}}

\newcommand{\cM}{\mathcal{M}}
\newcommand{\cN}{\mathcal{N}}
\newcommand{\cO}{\mathcal{O}}

\newcommand{\cV}{\mathcal{V}}
\newcommand{\cW}{\mathcal{W}}

\newcommand{\bC}{\mathbb{C}}

\newcommand{\bP}{\mathbb{P}}

\newcommand{\fg}{\mathfrak{g}}

\newcommand{\fp}{\mathfrak{p}}

\newcommand{\fs}{\mathfrak{s}}
\newcommand{\ft}{\mathfrak{t}}

\usepackage[Symbol]{upgreek}
\usepackage{bm}

\usepackage{calligra}
\DeclareMathAlphabet{\mathcalligra}{T1}{calligra}{m}{n}

\theoremstyle{plain}

  \theoremstyle{definition}

\providecommand{\examplename}{Example}
\providecommand{\theoremname}{Theorem}

\makeatletter
\g@addto@macro\bfseries{\boldmath}
\makeatother

\makeatletter
\newcommand*{\rom}[1]{\expandafter\@slowromancap\romannumeral #1@}
\makeatother

\title{Gluing gravitational blocks for AdS black holes}

\author[a]{Seyed Morteza Hosseini,}
\author[b]{Kiril Hristov,}
\author[c,d]{and Alberto Zaffaroni}
\affiliation[a]{Kavli IPMU (WPI), UTIAS, The University of Tokyo, Kashiwa, Chiba 277-8583, Japan}
\affiliation[b]{Institute for Nuclear Research and Nuclear Energy, Bulgarian Academy of Sciences, \\Tsarigradsko Chaussee 72, 1784 Sofia, Bulgaria}
\affiliation[c]{Dipartimento di Fisica, Universit\`a di Milano-Bicocca, I-20126 Milano, Italy}
\affiliation[d]{INFN, sezione di Milano - Bicocca, I-20126 Milano, Italy}
\emailAdd{morteza.hosseini@ipmu.jp}
\emailAdd{khristov@inrne.bas.bg}
\emailAdd{alberto.zaffaroni@mib.infn.it}

\preprint{IPMU19-0132}

\abstract{We provide a unifying entropy functional and an extremization principle for  black holes and black strings in AdS$_4\times S^7$ and AdS$_5\times S^5$ with arbitrary rotation and generic electric and magnetic charges. This is done by gluing gravitational blocks, basic building blocks that are directly inspired by the holomorphic blocks appearing in the factorization of supersymmetric partition functions in three and four dimensions. We also provide an explicit realization of the attractor mechanism by identifying the values of the scalar fields at the horizon with the critical points of the entropy functional. We give examples based on dyonic rotating  black holes with a twist in AdS$_4\times S^7$, rotating black strings in AdS$_5\times S^5$, dyonic Kerr-Newman black holes in AdS$_4\times S^7$  and Kerr-Newman black holes in AdS$_5\times S^5$. In particular, our entropy functional extends existing results by adding rotation to the twisted black holes in AdS$_4$ and by adding flavor magnetic charges for the Kerr-Newman black holes in AdS$_4$. We also discuss generalizations to higher-dimensional black objects.
}

\begin{document}

\setcounter{tocdepth}{2}
\maketitle

%
%

\date{Dated: \today}




\section{Introduction}
\label{sect:intro}

There has been some recent progress in the microscopical explanation of the entropy of BPS AdS black holes, initiated with the counting of microstates for static magnetically charged AdS$_4\times S^7$ black holes \cite{Benini:2015eyy} and continued, more recently, with partial counting for electrically charged and rotating black holes in AdS$_5\times S^5$ \cite{Cabo-Bizet:2018ehj,Choi:2018hmj,Benini:2018ywd}. These results have been extended to other compactifications and other dimensions. The microscopic counting is achieved by computing, using localization, the logarithm $\log Z(\nu^I)$ of the grand-canonical partition function of the holographically dual field theory, which corresponds either to the topologically twisted index or the superconformal one, and obtaining the entropy via a Legendre transform with respect to a set of chemical potentials $\nu^I$. The gravitational counterpart of this computation is usually encoded in an {\it attractor mechanism} in the spirit of \cite{Ferrara:1996dd,Ferrara:1995ih,Ooguri:2004zv,Sen:2005wa}. In this approach, the black hole entropy is obtained by extremizing an entropy functional  $\cI(\nu^I)$ with respect to the horizon value $\nu^I$ of a set of scalar fields and other modes. For example, the field theory computation for  AdS$_4$ black holes performed in \cite{Benini:2015eyy} perfectly matches with the attractor mechanism in $\cN=2$ gauged supergravity \cite{Cacciatori:2009iz,DallAgata:2010ejj}. Since not for all black holes the attractor mechanism has been studied and found in supergravity, it is often useful to write directly, using combined field theory and gravity intuition, an entropy functional $\cI(\nu^I)$ that reproduces the entropy of existing black holes. This approach was successfully used for electrically charged and rotating AdS$_5\times S^5$ black
holes in \cite{Hosseini:2017mds}, where it has been shown that the entropy functional has the remarkably simple form in terms of chemical potentials $\Delta^a$, $a = 1,2,3$, and $\omega_i$, $i=1,2$, conjugated, respectively, to the electric charges $Q_a$ and angular momenta $J_i$, 
\be\label{HHZ}
 \cI(\Delta_a,\omega_i)=  \ii \pi N^2 \, \frac{\Delta^1\Delta^2\Delta^3}{\omega_1\omega_2} + 2 \pi \ii \left (\sum_{a=1}^3 \Delta^a Q_a -  \sum_{i=1}^2 \omega_i J_i\right) ,
\ee
with the constraint $\Delta^1+\Delta^2+\Delta^3 +\omega_1+\omega_2 = 1$, where $N$ is the number of colors of the dual $\cN=4$ super Yang-Mills (SYM) theory. This result has been used in the later developments  \cite{Cabo-Bizet:2018ehj,Choi:2018hmj,Benini:2018ywd}. Entropy functional for other electrically charged and rotating black holes in diverse dimensions has been later found in \cite{Hosseini:2018dob,Choi:2018fdc} and, in some cases,  successfully compared to quantum field theory expectations, at least in particular limits. These entropy functionals can be also obtained by computing the zero-temperature limit of the on-shell action of a class of supersymmetric but nonextremal Euclidean black holes \cite{Cabo-Bizet:2018ehj,Cassani:2019mms}.  

In this paper we  provide a (field theory inspired) unifying entropy functional for spherical black holes and strings in AdS$_4\times S^7$ and AdS$_5\times S^5$ with arbitrary rotation and generic electric and magnetic charges. These include dyonic rotating  black holes with a twist  in AdS$_4\times S^7$  \cite{Cacciatori:2009iz,Katmadas:2014faa,Halmagyi:2014qza,Hristov:2018spe}, rotating black strings in AdS$_5\times S^5$ \cite{Benini:2013cda,Hosseini:2019lkt}, dyonic Kerr-Newman black holes in AdS$_4\times S^7$ \cite{Cvetic:2005zi,Hristov:2019mqp} and Kerr-Newman black holes in AdS$_5\times S^5$ \cite{Gutowski:2004ez,Kunduri:2006ek}. 
In order to give a unifying picture it is convenient to use a four-dimensional point of view.  All the above mentioned  black objects  can be dimensionally reduced to give four-dimensional rotating black hole solutions of an $\cN=2$ gauged supergravity coupled to vector multiplets that can be studied using the methods in \cite{Hristov:2018spe,Hristov:2019mqp}.  The relevant gauged supergravity arises as a consistent truncation of type IIB or M-theory and it is completely specified by a prepotential $\cF (X^\Lambda)$ and a set of gauging, or Fayet-Iliopoulos parameters, $\{g^\Lambda,g_\Lambda \}$.  
 
Our main result is the following. Consider a black hole with magnetic and electric charges encoded in the symplectic vector $\{ p^\Lambda, q_\Lambda \}$ and angular momentum $\cJ$.\footnote{The BPS conditions impose a linear constraint on the magnetic charges and some non-linear constraints among the remaining conserved quantities. These constraints are reflected in the constraints among chemical potentials.}  The corresponding  entropy functional 
is given by
\be\label{gluing}
 \cI ( p^\Lambda , \chi^\Lambda , \omega) \equiv  \frac{\pi}{4 G_{\text{N}}^{(4)}} \left (\sum_{\sigma = 1}^2 \cB \big(X^\Lambda_{(\sigma)} , \omega_{(\sigma)} \big) -2 \ii \chi^\Lambda q_\Lambda - 2 \omega \cJ \right )\, ,
\ee
where $\chi^\Lambda$ and $\omega$ are the chemical potentials conjugated to $q_\Lambda$ and $\cJ$, respectively. The entropy functional is obtained by  gluing a  quantity that we dub \emph{gravitational block}
\be
 \cB (X^\Lambda , \omega) \equiv - \frac{\cF(X^\Lambda)}{\omega} \, .
\ee
For black holes that are topologically twisted in four dimensions we use the \emph{$A$-gluing} 
\bea\label{Agluing}
 X^{\Lambda}_{(1)} & = \chi^\Lambda - \ii \omega \? p^\Lambda \, , \qquad && \omega_{(1)} = \omega \, , \\
 X^{\Lambda}_{(2)} & = \chi^\Lambda + \ii \omega \? p^\Lambda \, , && \omega_{(2)} = - \omega  \, .
\eea
while for the others we  use  the \emph{identity gluing} (\emph{id}-gluing) 
\bea\label{identitygluing}
 X^{\Lambda}_{(1)} & = \chi^\Lambda - \ii \omega \? p^\Lambda\, , \qquad && \omega_{(1)} = \omega \, , \\
 X^{\Lambda}_{(2)} & = \chi^\Lambda + \ii \omega \? p^\Lambda \, , && \omega_{(2)} = \omega  \, .
\eea
The functional $\cI$ must be extremized with respect to the chemical potentials $\chi^\Lambda$ and $\omega$ conjugated to $q_\Lambda$ and $\cJ$, respectively, and subject to a constraint that depends on the model.
Details of the model, type of gluing and constraint are explicitly given in the following table for all the above mentioned black holes:
\begin{table}[H]
  \begin{center}
       \label{tab:table1}
    \begin{tabular}{|c|c|c|c|}
    \hline
   \,  Black object & Gluing &Constraint & $\cF(X^\Lambda)$   \\
      \hline
    \emph{m}AdS$_4$ & $A$-gluing & $g_\Lambda \chi^\Lambda = 2$ & $2\ii \sqrt{X^0 X^1 X^2 X^3}$  \\
      \hline
   AdS$_5$ BS & $A$-gluing & $g_\Lambda \chi^\Lambda = 2$ &  $\displaystyle\frac{X^1 X^2 X^3}{X^0_{~}}$\mystrut   \\    
    \hline
      KN-AdS$_4$ & \emph{id}-gluing & $g_\Lambda \chi^\Lambda - \ii \omega = 2$ & $2 \ii \sqrt{X^0 X^1 X^2 X^3}$     \\
     \hline
      KN-AdS$_5$ & $A$-gluing  & $g_\Lambda \chi^\Lambda - \ii \omega \tanh (\delta) = 2$ &  $\displaystyle\frac{X^1 X^2 X^3}{X^0_{~}}$\mystrut  \\
      \hline
        \end{tabular}
     \caption{In this table,  \emph{m}AdS$_4$  refers to magnetically charged black holes in AdS$_4$ with a twist, BS$=$black strings and KN$=$Kerr-Newman. All the  black objects in AdS$_5$ are considered after dimensional reduction to four dimensions. The prepotential and gaugings can be read off from the existing consistent truncations of AdS$_4\times S^7$ and AdS$_5\times S^5$. The gaugings are purely electric. In suitable normalizations we can set  $g_\Lambda=1$ for AdS$_4$ black holes, $g_\Lambda=\{0,1,1,1\}$ for AdS$_5$ BS and  $g_\Lambda =\sqrt{2}\{\cosh (\delta),1,1,1\}$ for KN-AdS$_5$. The extra parameter $\delta$ appearing in KN-AdS$_5$ is an artifact of dimensional reduction and its role is explained in section \ref{sec:KNAdS5}. Notice that KN black holes have no twist in five dimensions but acquire one upon dimensional reduction to four.}
  \end{center}
\end{table}

\begin{figure}[t]
\label{fig:1:GB}
\begin{center}
 \begin{tikzpicture}[font=\footnotesize, scale=0.4]
 \draw ellipse (2 and 4);
  \draw (0,4) arc(90:270:4) ;
  \draw[fill=blue] (-4,0) circle(0.3);
  \node[left]  at (-4.2,0) {${\rm SP}$};
  \node[left]  at (-4.9,-2) {$\cF(X_{(1)}^\Lambda)$};
 \draw (8,4) arc(90:270:2 and 4) ;
 \draw[dashed] (8,-4) arc(-90:90:2 and 4) ;
  \draw(8,-4) arc(-90:90:4) ;
  \draw[fill=blue] (12,0) circle(0.3);  
 \node[right]  at (12.2,0) {${\rm NP}$};
 \node[right]  at (12.9,-2) {$\cF(X_{(2)}^\Lambda)$};
 \end{tikzpicture}
  \caption{Gluing gravitational blocks}\label{fig:1}
 \end{center}
 \end{figure}
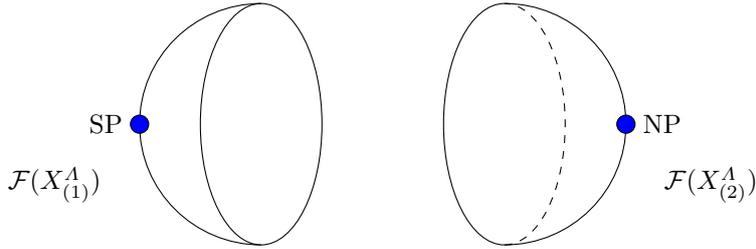

The previous construction gives an explicit realization of the {\it attractor mechanism}. The extremal value of the functional $\cI$ reproduces  the entropy of the black hole 
\be
 S_{\text{BH}} (p^\Lambda , q_\Lambda , \cJ) = \cI ( p^\Lambda, \chi^\Lambda , \omega) \Big|_{\text{crit.}} \, ,
\ee
and, as we will show, the extremal values of the quantities $X^\Lambda_{(1)}$ and  $X^\Lambda_{(2)}$ can be identified with the values of the supergravity sections $X^\Lambda$ at the South pole (SP) and the North pole (NP) of the sphere in the near horizon region (see Fig.\,\ref{fig:1:GB}). 
From this point of view, we can associate the two gravitational blocks entering in the gluing to the SP and NP of the sphere. The poles of the sphere are special because they are the two fixed points of the rotational symmetry.
This is in the spirit of previous formulations with rotation \cite{Astefanesei:2006dd,Hristov:2018spe}.

The entropy functional $\cI$ reproduces all the known results in the literature and generalizes them. In particular,  the functional with an $A$-gluing correctly reduces for $\omega=0$ to the standard attractor mechanism for static black holes with horizon geometry AdS$_2\times S^2$ \cite{Cacciatori:2009iz,DallAgata:2010ejj} for an arbitrary prepotential $\cF$. Indeed, in the limit $\omega \rightarrow 0$, \eqref{gluing} becomes
\be
 \cI_{\text{\emph{sm}AdS$_{4}$}} ( p^\Lambda , \chi^\Lambda) =  \ii \frac{\pi}{2 G_{\text{N}}^{(4)}} \left ( p^\Lambda F_\Lambda(\chi)  - q_\Lambda \chi^\Lambda \right ) ,
\ee
where, as usual, $F_\Lambda \equiv \partial_{\chi^\Lambda} \cF$. This is equivalent to the attractor mechanism \cite{Cacciatori:2009iz,DallAgata:2010ejj} 
\be
 S^{\text{\emph{sm}AdS$_{4}$}}_{\text{BH}} (p^\Lambda , q_\Lambda) = -\ii \frac{\pi}{G_{\text{N}}^{(4)}} \frac{p^\Lambda F_\Lambda - q_\Lambda X^\Lambda}{g^\Lambda F_\Lambda - g_\Lambda X^\Lambda} \, ,
\ee
once we choose  the convenient gauge $g_\Lambda X^\Lambda=2$ and we identify $X^\Lambda=\chi^\Lambda$.%
\footnote{Recall that the symplectic sections $X^\Lambda$ in supergravity are only defined up to rescaling. We will use this freedom often to find a convenient normalization for our quantities.
Recall also  that in our models the magnetic gaugings vanish, $g^\Lambda=0$.} The quantity $\cI$  also correctly reproduces the entropy functional \eqref{HHZ} for Kerr-Newman AdS$_5\times S^5$ black holes and the analogous one for  
Kerr-Newman AdS$_4\times S^7$ black holes, as we discuss in sections \ref{sec:KNAdS4} and \ref{sec:KNAdS5}.
Furthermore, our entropy functional \eqref{gluing} extends these known results  to the case of rotating twisted black holes in AdS$_4$ and to the case of Kerr-Newman black holes in AdS$_4$ with  flavor magnetic charges.
The entropy functional for rotating black strings in AdS$_5$ was already discussed in \cite{Hosseini:2019lkt} and inspired this investigation.

Our gravitational construction is inspired and closely related to the construction of three-dimensional supersymmetric partition functions by gluing holomorphic blocks \cite{Beem:2012mb}.%
\footnote{The idea of ``gluing'' or ``sewing'' building blocks to compose field theory observables has been put forward in different context also earlier, \cf\;\cite{Belavin:1984vu} and \cite{Nekrasov:2003vi} and references thereof. See also \cite{Pasquetti:2011fj,Hwang:2012jh,Imamura:2013qxa,Yoshida:2014ssa,Hwang:2015wna,Benini:2015noa,Nieri:2015yia,Gukov:2017kmk,Closset:2018ghr} for other related developments.}
In this context, most of the three-dimensional supersymmetric partition functions, and in particular the topologically twisted index and the superconformal one, can be written by gluing two holomorphic blocks according to the formula
\be
 \label{Zhol}
 Z(\Delta^\Lambda | \omega) = \sum_\alpha   B^\alpha ( \Delta^\Lambda_{(1)} | \omega_{(1)} ) B^\alpha ( \Delta^\Lambda_{(2)} | \omega_{(2)} ) \, .
\ee
Here, the holomorphic block, $B^\alpha(\Delta^\Lambda |\omega)$,  depends on the chemical potentials $\Delta^\Lambda$ for global symmetries and the equivariant parameter $\omega$, as well as
on a choice of Bethe vacuum for the two-dimensional theory obtained by reducing the theory on a circle. In applications to holography, we typically work in a saddle point approximation where  one Bethe vacuum 
dominates the sum \eqref{Zhol}. In this context, our gravitational blocks are holographically dual to holomorphic blocks in the Cardy limit (see \eg\;\cite[(2.22)]{Beem:2012mb} and \cite[(F.15)]{Closset:2018ghr}) 
\be\label{Bethehol}
 B^\alpha ( \Delta^\Lambda | \omega ) \underset{\omega \to 0}{\sim} \exp \bigg( \! -\frac{1}{\omega} \cW ( x^\alpha , \Delta^\Lambda ) \bigg) \, ,
\ee
where $\cW (x , \Delta^\Lambda )$ is the effective twisted superpotential of the two-dimensional theory and the Bethe vacua $x^\alpha$ are its critical points. It has been shown indeed in \cite{Hosseini:2016tor,Hosseini:2016cyf} that, upon the identification of $\Delta^\Lambda$ with $X^\Lambda$, and up to normalizations, the  twisted superpotential evaluated on the Bethe vacuum which is relevant for holography%
\footnote{The on-shell twisted superpotential of many three-dimensional $\cN=2$ Chern-Simons-matter gauge theories with holographic duals were computed in \cite{Hosseini:2016ume,Ray:2019lqb,Wearetookind:2019euv}.}
can be identified  with the supergravity prepotential
\bea
 \label{ident}
 & \cW ( x^\alpha , \Delta^\Lambda ) \Big|_{\text{BA}} \equiv \wt \cW (\Delta^\Lambda) =  \cF (X^\Lambda) \, ,
\eea
for all the theories that we discuss in this paper. The analogy can be pushed further. As discussed in section \ref{sec:mAdS4} and \ref{sec:BS}, the gravitational $A$-gluing \eqref{Agluing} precisely corresponds to the field theory gluing used for the topologically twisted index  \cite{Nieri:2015yia}. The identity gluing \eqref{identitygluing} is not exactly identical to the one used for the superconformal index \cite{Nieri:2015yia}. However, the identification \eqref{ident} is valid for a particular range of the complex variables  $\Delta^\Lambda$ that is not  always respected by the field theory gluing. We expect that, taking into account the necessary redefinitions, the physics of the two gluings is the same. It is not impossible that there is an alternative and more clever way of rewriting \eqref{identitygluing}. 

It would be interesting to derive our entropy functional from field theory. In particular, it is natural to conjecture that the large $N$ limit of the Legendre transform of the logarithm of the relevant topologically twisted index or superconformal index is given by the functional $\cI$, either in full generality or in suitable limits. Many partial checks of this already exist in the literature.%
\footnote{This has been checked at large $N$ in full generality for static \emph{m}AdS$_4$ black holes \cite{Benini:2015eyy,Benini:2016rke}, rotating black strings in AdS$_5$ \cite{Hosseini:2019lkt}, KN-AdS$_5$ black holes with equal angular momenta \cite{Benini:2018ywd} and at large $N$ but in the Cardy limit for general purely electric KN-AdS$_4$ and KN-AdS$_5$ \cite{Choi:2018hmj,Choi:2019zpz}. It is still not known if the large $N$ limit of the superconformal index of $\cN=4$ SYM reproduces the entropy functional, and therefore the entropy, in the case of KN-AdS$_5$ black holes with \emph{unequal} angular momenta.}
We see that our functional seems to arise from a saddle point approximation of \eqref{Zhol}  and it would be interesting to make this statement more precise. 
The $\omega\rightarrow 0$ limit of \eqref{Zhol} has been analyzed in \cite{Closset:2018ghr} and shown to reproduce the Bethe ansatz formula that has been used to derive the entropy functional for static black holes in \emph{m}AdS$_4$. Unfortunately, a field theory computation 
for rotating black holes in \emph{m}AdS$_4$ is still missing. 
On the other hand, the entropy functional for rotating black strings in AdS$_5$ has been derived explicitly from the topologically twisted index  in \cite{Hosseini:2019lkt} and it would be interesting to rederive the same result from  \eqref{Zhol}. Our result suggests that the very crude Cardy approximation \eqref{Bethehol} gives the right result also for finite $\omega$, at least  in the large $N$ limit. Finally, a similar but slightly different approach based on factorization of partition functions has been used to derive the entropy of Kerr-Newman AdS$_4$ black holes without magnetic charges in the Cardy limit in \cite{Choi:2019zpz}  and it would be interesting to extend it to other cases as well.%
\footnote{See also \cite{Nian:2019pxj} for a different approach.}

Our entropy functional can be generalized to black objects in six and seven  dimensions, including Kerr-Newman black holes \cite{Chow:2008ip,Cvetic:2005zi,Chow:2007ts} and magnetically charged twisted black objects \cite{Hosseini:2018usu,Benini:2013cda}   in the AdS$_6 \times_w S^4$ background of massive type IIA supergravity \cite{Brandhuber:1999np} and in AdS$_7\times S^4$. The structure of the higher-dimensional gravitational blocks is discussed in section  \ref{sec:higher}. 

As a final note, we observe that the entropy functional \eqref{gluing} is  strongly suggesting that some equivariant localization is at work in gravity. We will comment more on this in the discussion section.

This paper is organized as follows. In section \ref{sec:mAdS4} we discuss the general class of rotating  black holes in AdS$_4$ with non-vanishing magnetic charge for the R-symmetry found in \cite{Hristov:2018spe}. In section \ref{sec:BS} we discuss the  rotating black strings in AdS$_5$ found in \cite{Hosseini:2019lkt} and their reduction to four dimensions.  In section \ref{sec:KNAdS4} we discuss the case of Kerr-Newman black holes in AdS$_4$ with  magnetic charges for the flavor symmetries found in \cite{Hristov:2019mqp}. In section \ref{sec:KNAdS5}  we discuss the case of Kerr-Newman black holes in AdS$_5\times S^5$ \cite{Gutowski:2004ez,Kunduri:2006ek} and their reduction to four dimensions, generalizing \cite{Hosseini:2017mds}. For all these examples we show that the extremization of the entropy functional \eqref{gluing} reproduces the entropy and that the values of the sections $X^\Lambda$ at the NP and SP of the sphere are directly related to the gluing quantities \eqref{Agluing} and \eqref{identitygluing} evaluated at the critical point of $\cI$. In particular, this provides general analytical formulae for the critical point of the functional \eqref{gluing} that would be difficult to find with other methods. 
In section \ref{sec:higher} we discuss the natural generalization of our construction to higher dimensions. 
We conclude with discussion and outlook in section \ref{sect:Discussion}. Appendix \ref{app:A} contains some technical aspects of the relevant $\cN=2$ gauged supergravity and the black hole solutions of \cite{Hristov:2018spe,Hristov:2019mqp}.
Appendix \ref{app:B} contains an example of our techniques applied to asymptotically flat black holes in four dimensions. 

\section[Rotating black holes in \texorpdfstring{\emph{m}}{m}AdS\texorpdfstring{$_4$}{(4)}]{Rotating black holes in \emph{m}AdS$_4$}
\label{sec:mAdS4}

We first look at rotating black holes in AdS$_4$ with non-vanishing magnetic charge for the R-symmetry and near horizon AdS$_2 \times_w S^2$ \cite{Hristov:2018spe}.
The entropy of this class of black holes is supposed to be reproduced by the refined topologically twisted index of the holographically dual field theory on $S^2_\omega \times S^1$ \cite{Benini:2015noa}.
These solutions admit a static limit, originally found in \cite{Cacciatori:2009iz} in the purely magnetic case and in \cite{Katmadas:2014faa,Halmagyi:2014qza} in the dyonic case.
The non-vanishing magnetic charge means that the solutions are only asymptotically {\it locally} AdS$_4$, and the particular asymptotic spacetime was dubbed {\it magnetic} AdS$_4$ or just \emph{m}AdS$_4$ in \cite{Hristov:2011ye}.
This amounts to a partial topological $A$-twist on the two-sphere in the boundary field theory \cite{Hristov:2013spa}, so that some supersymmetry is preserved.

From the start we commit ourselves to the so-called \emph{magnetic} STU model of four-dimensional $\U(1)$ gauged supergravity.
It admits an embedding in the maximal $\SO(8)$ gauged supergravity in four dimensions and a further uplift on $S^7$ to eleven-dimensional supergravity \cite{Cvetic:1999xp}.
The dual field theory is ABJM \cite{Aharony:2008ug}, and in the absence of rotation this particular model provided the first successful microscopic counting for asymptotically AdS black holes \cite{Benini:2015eyy}.

The magnetic STU model is characterized by a prepotential
\be
 \cF (X^\Lambda) = 2 \ii \sqrt{X^0 X^1 X^2 X^3} \, ,
\ee
together with a purely electric gauging
\be
 G = \{g^{\Lambda}; g_{\Lambda} \} \, , \qquad g_\Lambda \equiv g \, , \qquad g^\Lambda = 0 \, .
\ee
We further choose to set $g=1$, thus fixing the AdS$_4$ scale $l^{2}_{\text{AdS}_4} = 1 / 2$.
See appendix \ref{app:A1} for a summary of the main features of four-dimensional ${\cal N}=2$ supergravity and the notations we employ for symplectic vectors and the quartic invariant $I_4$ we that will use in the following.

We are interested in the solution of \cite[sect.\,5]{Hristov:2018spe}, particularly in the near horizon geometry and attractor mechanism for the scalars. The full solution is characterized uniquely by the symplectic vector of gauging parameters $G$ introduced above, the conserved angular momentum $\cJ$, and the symplectic vector of conserved electromagnetic charges
\be
 \Gamma = \{ p^\Lambda ; q_\Lambda  \} \, .
\ee
The twisting condition imposes the following relation between the magnetic charges,
\be\label{twist1}
 \sum_{\Lambda = 0}^{3} p^\Lambda = -1 \, .
\ee 

All the relevant quantities can be expressed  in terms of  the quartic invariant
\bea\label{eq:mSTUI4}
I_4(\Gamma)=&\? - (p^0 q_0 - p^i q_i)^2 + 4\,q_0 q_1 q_2 q_3 + 4\, p^0 p^1 p^2 p^3 \\
                              &\?   + 4\? (p^1 p^2 q_1 q_2 + p^1 p^3 q_1 q_3 + p^2 p^3 q_2 q_3) \, .
\eea
The full solution for the metric, gauge fields and scalars is summarized in appendix \ref{app:A2}
and depends on the symplectic vector $\cH_0$ and the extra parameter ${\rm j}$, which are then related to $\cJ, \Gamma, G$ via the attractor equations \eqref{eq:attr-twisted} and \eqref{eq:bigJ-twisted}.
The solution for $\cH_0$ and ${\rm j}$ can be explicitly found in \cite[sect.\,5]{Hristov:2018spe}.

The main quantity of interest, the Bekenstein-Hawking entropy, reads
\be\label{BHmAdS4}
 S^{\text{\emph{m}AdS$_{4}$}}_{\text{BH}} (p , q,  \cJ) = \frac{\pi}{2 \sqrt{2} G^{(4)}_{\text{N}}} \sqrt{F_2 + \sqrt{\Theta - 16 \cJ^2}} \, ,
\ee
where we defined
\be
 F_2 \equiv \frac14 I_4(\Gamma, \Gamma, G, G) \, , \qquad \Theta \equiv (F_2)^2 - 16\? I_4 (\Gamma) \, ,
\ee
in agreement with $\Theta$ and $F_2$ in \cite[app.\,A]{Benini:2015eyy}.
In the case of vanishing electric charges, $\Gamma = \{p^{\Lambda}; 0\}$,
\be
 F_2 = \sum_{\Lambda < \Omega} p^\Lambda p^\Omega - \sum_{\Lambda = 0}^3 ( p^\Lambda )^2 \, , \qquad \Theta = (F_2)^2 - 64\? p^0 p^1 p^2 p^3 \, ,
\ee
while in general when $q_\Lambda \neq 0$ the explicit expressions are rather long and not particularly illuminating.
The chemical potential $w$ conjugate to the angular momentum \cite{Hristov:2018spe} is given by
\be
 \label{w:mAdS4}
 w = \frac{2 \sqrt{2}\? {\cal J}}{\sqrt{\Theta - 16\? {\cal J}^2}\? \sqrt{F_2 + \sqrt{\Theta - 16 \cJ^2}} } \, .
\ee
We will first look at the general attractor mechanism predicted from the gluing procedure. After that we will describe the solution for the symplectic sections $X^\Lambda$ at the near horizon,
initially in a simplified setting with reduced number of charges and then in general.

\subsection{Attractor mechanism}

From here on we use a ``field theoretical normalization'' for the magnetic charges $\fp^\Lambda = - 2 p^\Lambda$, that allows for a better comparison with existing literature \cite{Benini:2015eyy,Benini:2016rke}.
The twisting condition \eqref{twist1} becomes
\be
 \sum_{\Lambda=0}^3 \fp^\Lambda =2 \, ,
\ee
corresponding to the fact that the superpotential of ABJM \cite{Aharony:2008ug} has R-charge two.

In a model with a topological twist, we need to use the $A$-gluing \eqref{Agluing} that, in the new normalization, reads
\bea
 X^{\Lambda}_{(1)} & = \chi^\Lambda + \ii \frac{\omega}{2} \? \fp^\Lambda \, , \qquad && \omega_{(1)} = \omega \, , \\
 X^{\Lambda}_{(2)} & = \chi^\Lambda - \ii \frac{\omega}{2} \? \fp^\Lambda \, , && \omega_{(2)} = - \omega  \, ,
\eea
and the constraint on chemical potentials is
\be
 \label{const:mAdS4}
 \sum_{\Lambda=0}^3 \chi^\Lambda =2 \, ,
\ee
where we have set $g_\Lambda=1$.

 The entropy functional is then  given by \eqref{gluing}:
\be
 \label{entropy:functional:mAdS}
 \cI_{\text{\emph{m}AdS$_{4}$}} ( \fp^\Lambda , \chi^\Lambda , \omega ) \equiv  \frac{\pi}{4 G_{\text{N}}^{(4)}}
 \left( \cE_{\text{\emph{m}AdS$_{4}$}} ( \fp^\Lambda , \chi^\Lambda , \omega ) - 2 \ii \chi^\Lambda q_\Lambda - 2 \omega \cJ \right )
 + \lambda \bigg( \sum_{\Lambda = 0}^3 \chi^\Lambda - 2 \bigg) \, ,
\ee
where 
\be
 \cE_{\text{\emph{m}AdS$_{4}$}} ( \fp^\Lambda , \chi^\Lambda , \omega ) = - \frac{2 \ii}{\omega}  \left (  \sqrt{X_{(1)}^0 X_{(1)}^1 X_{(1)}^2 X_{(1)}^3} -\sqrt{X_{(2)}^0 X_{(2)}^1 X_{(2)}^2 X_{(2)}^3} \right )  ,
\ee
and we introduced a Lagrange multiplier to enforce the constraint on chemical potentials \eqref{const:mAdS4}.
Note that due to the presence of a square root there are sign ambiguities to take into account when performing the extremization. They correspond to different branches in the parameter space of the black hole solution. Notice that not all solutions that lead to a positive value for $\cI$ correspond to regular black holes. One should also check that there are no other singularities in the metric and this may restrict the range of the allowed conserved charges. This analysis can be only done case by case.

We first state the general result. The attractor mechanism works as follows. The values of the sections at the SP $(\theta = 0)$ and the NP $(\theta = \pi)$ of the sphere are  given by
\be\label{eq:mads4sections}
 X^\Lambda_{\text{SP, NP}} =\frac{\ii}{2} \Big( \bar \chi^\Lambda \pm \ii \frac{\bar \omega}{2} \fp^\Lambda \Big) \, , \qquad \Lambda = 0 , \ldots, 3 \, ,
\ee
where $\bar \chi$ and $\bar \omega$
are the critical points of the functional \eqref{entropy:functional:mAdS}. We see that the values of the sections can be identified with the critical values of the gluing quantities  $X^{\Lambda}_{(\sigma)}$ as 
\be 
X^\Lambda_{\text{SP}} =\frac{\ii}{2} \,  X^{\Lambda}_{(1)} \, \Big|_{\text{crit.}} \, ,   \qquad\qquad X^\Lambda_{\text{NP}} =\frac{\ii}{2} \,  X^{\Lambda}_{(2)} \, \Big|_{\text{crit.}} \, , \qquad \Lambda = 0 , \ldots, 3  \, . 
\ee 
Moreover, we find that
\be  S^{\text{\emph{m}AdS$_{4}$}}_{\text{BH}} (p^\Lambda , q_\Lambda, \cJ ) =  \cI_{\text{\emph{m}AdS$_{4}$}} ( \fp^\Lambda , \bar \chi^\Lambda , \bar \omega ) \, . \ee

The entropy functional  \eqref{entropy:functional:mAdS} generalizes the known result for static black holes, $\cJ = 0$, by taking the limit $\omega \rightarrow 0$,\footnote{Here we have taken the negative determination for the square root that is the one leading to regular black hole solutions \cite{Benini:2015eyy}. We inherit from \cite{Hristov:2018spe} an unfortunate choice of sign for the prepotential that leads to ambiguities in the comparison with the literature and we apologize to the reader for that.}
\bea
 \cE_{\text{\emph{sm}AdS$_{4}$}} ( \fp^\Lambda , \chi^\Lambda) &= - \sqrt{\chi^0 \chi^1 \chi^2 \chi^3} \sum_{\Lambda = 0}^3 \frac{\fp^\Lambda}{\chi^\Lambda}
 \, , \\
 S^{\text{\emph{sm}AdS$_{4}$}}_{\text{BH}} (p^\Lambda , q_\Lambda) &=  \cI_{\text{\emph{sm}AdS$_{4}$}} ( \fp^\Lambda , \chi^\Lambda) \Big|_{\text{crit.}} \, . 
\eea
This is precisely the result obtained in \cite{Benini:2015eyy,Benini:2016rke} upon identifying the variables $\Delta^\Lambda$ used  in \cite{Benini:2015eyy,Benini:2016rke}  with $\pi \chi^\Lambda$. In order to compare with field theory one also needs to use 
\be
 \label{AdS4:CFT3:dict}
 \frac{1}{G_{\text{N}}^{(4)}} = \frac{4 \sqrt{2}}{3} N^{3/2} \, .
\ee

\subsection[The purely magnetic \texorpdfstring{T$^3$}{T**3} model]{The purely magnetic T$^3$ model}

The T$^3$ model is obtained by setting
\be
 \label{three:eq:sections}
 \chi^{1,2,3} = \chi \, , \qquad \chi^0 = 2 - 3 \chi \, ,
\ee
and, similarly, for the magnetic fluxes
\be
 \label{three:eq:fluxes}
 \fp^{1,2,3} = \fp \, , \qquad \fp^0 = 2 - 3 \fp \, ,
\ee
while here for simplicity we set the electric charges to zero.
Therefore,
\be
 F_2 = - ( 1 - 6 \fp + 6 \fp^2 ) \, , \qquad \Theta = ( 1 - 2 \fp )^3 ( 1 - 6 \fp ) \, .
\ee
The values of the sections $X^0 (\theta)$ and $X^1 (\theta) = X^2 (\theta) = X^3 (\theta)$ at the horizon can be computed using \eqref{horsec} and when specified to the NP ($\theta = \pi$) and SP ($\theta = 0$) read
\be
 \begin{aligned}
  & X^\Lambda_{\text{SP}} - X^\Lambda_{\text{NP}} = w \fp^\Lambda \, , \qquad \Lambda = 0 , \ldots, 3 \, , \\
  & X^0_{\text{SP}} + X^0_{\text{NP}} = \frac{\ii}{2\? \sqrt{\Theta - 16\? \cJ^2}}\? \left( 3 - 12 \fp (1 - \fp) + \sqrt{\Theta - 16\? \cJ^2} \right)  , \\
  & X^1_{\text{SP}} + X^1_{\text{NP}} = \frac{\ii}{2\? \sqrt{\Theta - 16\? \cJ^2}}\? \left( -1 + 4 \fp (1 - \fp) + \sqrt{\Theta - 16\? \cJ^2} \right) , 
 \end{aligned}
\ee
where we need to take $\fp < 0$  in order to find regular black hole solutions, and the angular momentum is constrained in the range $|\cJ| < \sqrt{\Theta} / 4$.%
\footnote{More generally, in the full STU model with vanishing electric charges, the regions of positivity for $\fp^{1,2,3}$ where regular black holes exist were determined in \cite[app.\,A]{Benini:2015eyy} and the rotation is bounded above, $|\cJ| < \sqrt{\Theta} / 4$. In the most general case with electric charges one again requires positive scalars and $\Theta > 0$ with the same bound on $\cJ$, but due to the big number of free parameters the regions of positivity are much harder to determine and we have not pursued this here.} 

We have then checked numerically  that the values of the sections at the SP and the NP of the sphere are  given by
\be
 X^\Lambda_{\text{SP, NP}} =\frac{\ii}{2} \Big( \bar \chi^\Lambda \pm \ii \frac{\bar \omega}{2} \fp^\Lambda \Big) \, , \qquad \Lambda = 0 , \ldots, 3 \, ,
\ee
where $\bar \chi$ and $\bar \omega = - 2 w$
are the critical points of the functional $\cI_{\text{\emph{m}AdS$_{4}$}} (\fp , \chi , \omega)$ in  \eqref{entropy:functional:mAdS}. We have also checked that the critical value of the entropy functional reproduces the entropy \eqref{BHmAdS4} of the rotating black holes 
\be
 S^{\text{\emph{m}AdS$_{4}$}}_{\text{BH}} ( p^\Lambda , q_\Lambda, \cJ ) =  \cI_{\text{\emph{m}AdS$_{4}$}} ( \fp^\Lambda , \bar \chi^\Lambda , \bar \omega )  \, ,
\ee
where again, just as in the static case, one has to take the negative determination for the square root in $\cI_{\text{\emph{m}AdS$_{4}$}}$.

Notice the supergravity attractor mechanism provides general \emph{analytic} formulae for the critical point of the functional \eqref{entropy:functional:mAdS} that are very difficult to find by a direct extremization. 

\subsection{The general case}

In the general case there are at most six free parameters $\fp^{1,2,3}, \cJ, q_{1,2}$, since the electric charges are constrained by the requirement that there are no NUT charges,  see \eqref{eq:twistedconstraint}. One needs indeed to ensure that
\be
	I_4 (G, \Gamma, \Gamma, \Gamma) = I_4 (\Gamma, G, G, G) = 0\ .
\ee
The second constraint gives a linear constraint that can be easily solved by $q_0 = -(q_1+q_2+q_3)$ while  the first constraint gives a more complicated cubic relation among charges. 

The general expressions in the case of arbitrary charges are much more involved, but one can check numerically that  the black hole entropy is still given by the critical value of $\cI_{\text{\emph{m}AdS$_{4}$}} ( \fp^\Lambda , \chi^\Lambda , \omega )$. \eqref{eq:mads4sections} holds in full generality as well.

 In the static case $\cJ=0$ there is a microscopic counting of the entropy using field theory methods that  identifies $\frac{\pi}{4 G_{\text{N}}^{(4)}}  \cE( \fp^\Lambda , \bar \chi^\Lambda , 0)$ with the logarithm of the topologically twisted index of ABJM   \cite{Benini:2015eyy,Benini:2016rke}.    An analogous computation for rotating black holes would involve the \emph{refined} topologically twisted index defined in \cite{Benini:2015noa} and is still missing. As already mentioned in the introduction, our result suggests that the Cardy approximation \eqref{Bethehol} is actually exact and gives the right result also for finite $\omega$, at least  in the large $N$ limit. 

\section[Rotating black strings in AdS\texorpdfstring{$_5$}{(5)}]{Rotating black strings in AdS$_5$}
\label{sec:BS}

Our next example comes from the recently spelled out solutions of rotating AdS$_5$ black strings with near horizon BTZ$\times_w S^2$ \cite{Hosseini:2019lkt}.
Just like our previous example, the solutions we discuss here preserve supersymmetry by a twist on $S^2$ and admit a static limit.
The static solutions, with only magnetic charges, were originally found and understood as holographic RG flows across dimensions in \cite{Benini:2013cda}.
In particular, we have a flow from a UV four-dimensional $\cN=1$ theory compactified on $S^2$ to a two-dimensional $(0,2)$ theory in the IR whose exact central charges can be obtained employing $c$-extremization \cite{Benini:2012cz,Benini:2013cda}.
Due to the 4D/5D connection \cite{Gaiotto:2005gf,Behrndt:2005he} between solutions in gauged supergravity,
one can also understand these AdS$_5$ black strings as four-dimensional black holes in an asymptotically runaway spacetime \cite{Hristov:2014eza} and exploiting the relation also add electric charges \cite{Hristov:2014hza}.
Following \cite{Hosseini:2016cyf}, the \emph{refined} topologically twisted indices \cite{Benini:2015noa} of $\cN=1$ quiver gauge theories on $S^2_\omega \times T^2$ dual to rotating AdS$_5$ strings
were recently computed in \cite{Hosseini:2019lkt} in the Cardy limit giving a microscopic derivation of the entropy of this class of solutions.

Consider the \emph{electric} STU model in four dimensions. It is related to the five-dimensional gauged STU model via the 4D/5D connection
and thus it also admits an embedding in maximal $\SO(6)$ gauged supergravity in five dimensions. A further uplift on $S^5$ gives a solution of type IIB supergravity in ten dimensions \cite{Cvetic:1999xp}.
The holographically dual field theory is therefore $\SU(N)$ $\cN=4$ SYM.

The electric STU model is characterized by a prepotential
\be
 \cF (X^\Lambda) = \frac{X^1 X^2 X^3}{X^0} \, ,
\ee
and we have the purely electric gauging coming from the Kaluza-Klein reduction of the five-dimensional theory
\be
 G = \{g^{\Lambda}; 0 , g_{i} \} \, , \qquad g_i \equiv g \, , \qquad g^\Lambda = 0 \, .
\ee
We set $g=1$ for simplicity.\footnote{This choice amounts to setting the five-dimensional gauge coupling constant to $g_{(5)} = 1/\sqrt{2}$.} The symplectic vector of electromagnetic charges in this case reads
\be
 \Gamma = \{ 0 , p^i ; q_\Lambda  \} \, ,
\ee
with the twisting condition
\be
	\sum_{i = 1}^{3} p^i = -1 \, .
\ee
The condition $p^0 = 0$ stems from the fact that a compactification down to four dimensions along the length of the black string does \emph{not} introduce any magnetic charge.

The quartic invariant for the electric STU model can be written as
\bea
\label{eq:I4electricstu}
  I_4 (\Gamma) =& 4\? q_0 p^1 p^2 p^3 - \sum_{i = 1}^3 (p^i q_i)^2  + 2 \sum_{i<j}^3 q_i p^i q_j p^j\\
&- p^0 \bigg( 4\? q_1 q_2 q_3 + p^0 (q_0)^2 + 2\? q_0\? \sum_{i=1}^3 p^i q_i \bigg) \, , 
\eea
where the second row vanishes in the present case.
Again, the solution can be completely described in terms of ${\rm j}$ and $\cH_0$ that are uniquely fixed by the symplectic vectors $G, \Gamma$, as well as the angular momentum $\cJ$ as summarized in appendix \ref{app:A2}.
The explicit solution for ${\rm j}$ and $\cH_0$ can be found in \cite{Hosseini:2019lkt}.

In order to have a regular solution with a vanishing NUT charge we further require that, see \eqref{eq:twistedconstraint},
\be
 q_3 = \frac{p^1 ( p^1 - p^2 - p^3 ) q_1 - p^2 ( p^1 - p^2 + p^3 ) q_2}{p^3 ( p^1 + p^2 - p^3 )} \, .
\ee
We then find the following expression for the Bekenstein-Hawking entropy of the four-dimensional black holes,
\be
 S^{\text{AdS$_{5}$\,BS}}_{\text{BH}} \left( \fp^i , q_\Lambda , \cJ \right) = \frac{\pi}{G^{(4)}_{\text{N}}}\? \sqrt{\frac{- I_4(\Gamma)- {\cal J}^2}{\Theta}} \, ,
\ee
where we used
\be
\begin{aligned}
  & \Pi = ( - p^1 + p^2 + p^3) ( p^1 - p^2 + p^3 ) ( p^1 + p^2 - p^3 ) \, , \\
& \Theta = \sum_{i=1}^3 (p^i)^2 -  2 \sum_{i < j}^3 p^i p^j \, , \qquad  \Theta - \Pi = 8\? p^1 p^2 p^3 \, .
 \end{aligned}
\ee
The chemical potential conjugate to the angular momentum is also given by \eqref{eq:twsitedchempotJ},
\be
 w = -\frac{{\cal J}}{\sqrt{\Theta (- I_4(\Gamma) - {\cal J}^2)}} \, .
\ee
Finally, the sections in the near horizon region are found to be
\bea
 X^0 (\theta) & = 2\? \frac{p^1 p^2 p^3}{\sqrt{\Theta \? \Xi (\theta)}} \, , \\
 X^i (\theta) & = \frac{ p^i {\cal J} \cos (\theta)}{\sqrt{\Theta \? \Xi (\theta)}}
 + \ii\? \frac{ p^i ( 2\? p^i + 1 )}{\Theta}
 +2\? \frac{p^1 p^2 p^3}{( 2\? p^i +  1 ) \sqrt{\Theta\? \Xi (\theta)}} \bigg( \! q_i - \sum_{i=1}^3 q_i \! \bigg) \, , \quad i = 1, 2 , 3 \, ,
\eea
where we defined
\be
	  \Xi (\theta) \equiv (- I_4(\Gamma) - {\cal J}^2) + \frac{\Theta}{\Pi} {\cal J}^2 \sin^2 (\theta) \, .
\ee

\subsection{Attractor mechanism}
Since the theory is still topologically twisted in four dimensions, we need to use 
again the $A$-gluing \eqref{Agluing}:
\bea
 X^{\Lambda}_{(1)} & = \chi^\Lambda + \ii \frac{\omega}{2} \? \fp^\Lambda \, , \qquad && \omega_{(1)} = \omega \, , \\
 X^{\Lambda}_{(2)} & = \chi^\Lambda - \ii \frac{\omega}{2} \? \fp^\Lambda \, , && \omega_{(2)} = - \omega \, .
\eea
where we used  the ``field theory'' magnetic charges $\fp^\Lambda = - 2 p^\Lambda$.
There are only three non-vanishing magnetic charges and they satisfy
\be \sum_{i=1}^3 \fp^i =2 \, ,\ee
corresponding to the fact that the superpotential of $\cN=4$ super Yang-Mills (SYM) has R-charge two.

The entropy functional \eqref{gluing} is  given by
\be\label{BSI}
 \cI_{\text{AdS$_{5}$\,BS}} ( \fp^i , \chi^\Lambda , \omega ) \equiv  \frac{\pi}{4 G_{\text{N}}^{(4)}}
 \left( \cE_{\text{AdS$_{5}$\,BS}}( \fp^i , \chi^\Lambda , \omega ) - 2 \ii \chi^\Lambda q_\Lambda - 2 \omega \cJ \right )
 + \lambda \bigg( \sum_{i = 1}^3 \chi^i - 2 \bigg) \, ,
\ee
where
\be\label{BSI2}
 \cE_{\text{AdS$_{5}$\,BS}} ( \fp^i , \chi^\Lambda , \omega) = - \frac{\ii}{\chi^0} \left( \chi^1 \chi^2 \fp^3 + \chi^3 \chi^1 \fp^2 + \chi^2 \chi^3 \fp^1 - \frac{\omega^2}{4} \fp^1 \fp^2 \fp^3 \right) .
\ee
Here, we introduced the Lagrange multiplier $\lambda$ to enforce the constraint on the chemical potentials.
Notice that we can also write 
\be
 \cE_{\text{AdS$_{5}$\,BS}} ( \fp^i , \chi^\Lambda , \omega) = - \ii \sum_{i=1}^3 \fp^i \frac{\partial \cF(\chi) } {\partial \chi^i}  + \ii \frac{\omega^2}{24} \sum_{i,j,k=1}^3 \fp^i \fp^j \fp^k   \frac{\partial^3 \cF(\chi) } {\partial \chi^i\partial \chi^j\partial \chi^k} \, .
\ee
As a difference with the \emph{m}AdS$_4$ black holes of section \ref{sec:mAdS4}, here the Taylor series expansion  of the entropy functional \eqref{gluing} truncates at order $\cO(\omega^2)$ since the prepotential is cubic in the
variables $X^1$, $X^2$ and $X^3$ associated with nonzero magnetic charges.  

As it has already been checked in \cite{Hosseini:2019lkt}, the attractor mechanism works perfectly. The  values of the sections at the SP $(\theta = 0)$ and the NP $(\theta = \pi)$ are given by
\be
 X^\Lambda_{\text{SP, NP}} =\frac{\ii}{2} \Big( \bar \chi^\Lambda \pm \ii \frac{\bar \omega}{2} \fp^\Lambda \Big) \, , \qquad \Lambda = 0 , \ldots, 3 \, ,
\ee
where $\bar \chi$ and $\bar \omega = - 2 w$
are the critical points of the functional $\cI_{\text{AdS$_{5}$\,BS}} ( \fp^i , \chi^\Lambda , \omega )$. Moreover,
\be
 S^{\text{AdS$_{5}$\,BS}}_{\text{BH}} (p^i , q_\Lambda , \cJ) = \cI_{\text{AdS$_{5}$\,BS}} ( \fp^i , \bar \chi^\Lambda , \bar \omega) \, .
\ee
In order to compare with \cite{Hosseini:2019lkt} we need the redefinitions
\bea
 & \beta = - \frac{\ii \pi}{2} \chi^0 \, , \qquad &&\omega_{\text{there}} = \ii \pi \omega_{\text{here}} \, , \\
 & \Delta_i = \pi \chi^i \, , && i = 1 , 2 , 3 \, , \\
 & e_0 = \frac{1}{2 \sqrt{2} G^{(4)}_{\text{N}}} q_0 \, , && J = - \frac{1}{2 G^{(4)}_{\text{N}}} \cJ \, , \\
 & e_i = \frac{1}{2 \sqrt{2} g_{(5)} G^{(4)}_{\text{N}}} q_i \, , && i = 1 , 2 , 3 \, .
\eea
We note also the holographic relations
\be
 G^{(4)}_{\text{N}} = \frac{1}{2 \pi} G^{(5)}_{\text{N}} \, , \qquad  \frac{1}{g_{(5)}^3 G^{(5)}_{\text{N}}} = \frac{2}{\pi} N^{2} \, .
\ee

In  \cite{Hosseini:2019lkt} the functional $\cE_{\text{AdS$_{5}$\,BS}}$ \eqref{BSI} was explicitly derived as the logarithm of the refined topologically twisted index of $\cN=4$ SYM in the Cardy limit, thus providing a microscopic explanation of the entropy
of the four-dimensional black holes discussed in this section. 

\section[Kerr-Newman-AdS\texorpdfstring{$_4$}{(4)}]{Kerr-Newman-AdS$_{4}$}
\label{sec:KNAdS4}

Now let us we focus on the \emph{dyonic} Kerr-Newman black holes in AdS$_4$ with vanishing magnetic charge for the R-symmetry \cite{Hristov:2019mqp},
whose entropy is supposed to be reproduced by the generalized superconformal index of the holographically dual field theory on $S^2 \times S^1$ \cite{Kapustin:2011jm}.
Due to the vanishing magnetic flux for the R-symmetry, the asymptotic spacetime in this case is \emph{global} AdS$_4$ and thus full superconformal symmetry is preserved at the boundary.
However, we allow for magnetic fluxes for the extra flavor symmetries that are coming from the additional $\U(1)$ vector multiplets in the supergravity model. These additional fluxes break half of the supersymmetries at the boundary, in accordance with the generalized superconformal index.

We again consider the magnetic STU model with an uplift on $S^7$ to eleven dimensions.
The model is specified by the prepotential
\be
 \cF (X^\Lambda) = 2 \ii \sqrt{X^0 X^1 X^2 X^3} \, ,
\ee
and the purely electric gauging
\be
 G = \{g^{\Lambda}; g_{\Lambda} \} \, , \qquad g_\Lambda \equiv g \, , \qquad g^\Lambda = 0 \, .
\ee
We again set $g = 1$, fixing the AdS$_4$ length scale $l^{2}_{\text{AdS}_4} = 1 / 2$. The charge vector reads
\be
 \Gamma = \left\{ p^\Lambda ; q_\Lambda \right\} \, ,
\ee
with the same quartic invariant as in \eqref{eq:mSTUI4} and with the constraint that the R-symmetry magnetic flux vanishes,
\be
 \label{P:constraint:KNAdS4}
  \sum_{\Lambda = 0}^3 p^\Lambda = 0 \, .
\ee
The equations governing the solution, and consequently the solutions themselves, are much more involved in this case than for the twisted cases of the previous two sections. However, conceptually one again finds that the solution is entirely fixed by the symplectic vectors $G$ and $\Gamma$. Note that the angular momentum $\cJ$ in this case is never vanishing and is also uniquely fixed in terms of the electromagnetic charges. The near horizon solution is summarized in appendix \ref{app:A3} and fixed in terms of the symplectic vector $\cC$, which by the algebraic attractor equations depends on the charges, see \eqref{attr:eq:KNAdS4}. Unfortunately, in this case we cannot present in full generality a formula that gives $\cC$ in terms of $G$ and $\Gamma$, but one can always write down a complete solution in terms of auxiliary parameters entering $\cC$ and then express all physical quantities in terms of them. 

A very general solution, corresponding to four independent electric and one independent magnetic charge, can be written by the following parameterization of the vector $\cC$,
\be
\label{eq:C-parameterization}
 \cC = \left\{ - \alpha , \alpha , - \alpha \? \frac{\beta_0 - \beta_1}{\beta_2 - \beta_3} , \alpha \? \frac{\beta_0 - \beta_1}{\beta_2 - \beta_3} ; \beta_\Lambda \right\} .
\ee
This form of $\cC$ already satisfies the regularity constraints, see \eqref{eq:KNregularity}, and contains five out of the maximally allowed six independent parameters satisfying these conditions. It contains the purely electric KN-AdS$_4$ in the general STU model \cite{Cvetic:2005zi,Hristov:2019mqp}, the general dyonic $X^0 X^1$ model \cite{Hristov:2019mqp}, as well as more general solutions. 

The values of the conserved charges, the sections and the macroscopic entropy can be determined in terms of the parameters $\alpha, \beta_\Lambda$ in \eqref{eq:C-parameterization} via the formulae given in appendix \ref{app:A3}.
We first turn to the gluing formula and then check explicitly that it correctly predicts the attractor mechanism in several special cases of particular interest, therefore showing the general validity of the gluing procedure.
We choose not to present explicitly the most general allowed solution for $\cC$ since expressions soon become very cumbersome.

\subsection{Attractor mechanism}

Since there is no topological twist, we use the identity gluing \eqref{identitygluing}:
\bea
 X^{\Lambda}_{(1)} & = \chi^\Lambda + \ii \frac{\omega}{2} \? \fp^\Lambda \, , \qquad && \omega_{(1)} = \omega \, , \\
 X^{\Lambda}_{(2)} & = \chi^\Lambda - \ii \frac{\omega}{2} \? \fp^\Lambda \, , && \omega_{(2)} = \omega  \, ,
\eea
where we use again the notation $\fp^\Lambda = - 2 p^\Lambda$. The entropy functional \eqref{gluing} is given by
\be\label{eq:entropyfunctionalnotwist}
 \cI_{\text{KN-AdS$_{4}$}} ( \fp^\Lambda , \chi^\Lambda , \omega ) =  \frac{\pi}{4 G_{\text{N}}^{(4)}} \left (\cE_{\text{KN-AdS$_{4}$}} ( \fp^\Lambda , \chi^\Lambda , \omega )
 - 2 \ii \chi^\Lambda q_\Lambda - 2 \omega \cJ \right )
 - \lambda \bigg( \sum_{\Lambda = 0}^3 \chi^\Lambda - 2 - \ii \omega \bigg) \, ,
\ee
where 
\be\label{eq:entropyfunctionalnotwist2}
 \cE_{\text{KN-AdS$_{4}$}} ( \fp^\Lambda , \chi^\Lambda , \omega ) = - \frac{2 \ii}{\omega} \left( \sqrt{X_{(1)}^0 X_{(1)}^1 X_{(1)}^2 X_{(1)}^3} + \sqrt{X_{(2)}^0 X_{(2)}^1 X_{(2)}^2 X_{(2)}^3} \right) ,
\ee
and, as usual, we have introduced the constraint $\sum_{\Lambda = 0}^3 \chi^\Lambda  - \ii \omega=2$ through a Lagrange multiplier.
As in section \ref{sec:mAdS4}, due to the presence of a square root there are sign ambiguities when performing the extremization.
They correspond to different branches in the range of parameters, \ie\;conserved charges, for the black hole solution.  

The attractor mechanism in this case  works as follows.   For a suitable choice of determination of the square root in \eqref{eq:entropyfunctionalnotwist2}, the values of the sections at the SP $(\theta = 0)$ and the NP $(\theta = \pi)$ are related to the critical points
of the functional $\cI_{\text{KN-AdS$_{4}$}} ( \fp^\Lambda , \chi^\Lambda , \omega )$ by
\be\label{eq:nontwistedsectionscritical}
 \big( X^\Lambda_{\text{SP}} \big)^* = - \frac{\ii}{2} \left( \bar\chi^\Lambda + \ii \frac{\bar\omega}{2} \fp^\Lambda \right) \, , \qquad
 X^\Lambda_{\text{NP}} = - \frac{\ii}{2} \left( \bar\chi^\Lambda - \ii \frac{\bar\omega}{2} \fp^\Lambda \right) \, ,
\ee
for $\Lambda = 0, \ldots, 3$. Notice that here the values of the sections can be identified with the critical values of the gluing quantities  $X^{\Lambda}_{(\sigma)}$ up to a complex conjugate
\be 
\big (X^\Lambda_{\text{SP}}\big)^* = -\frac{\ii}{2} \,  X^{\Lambda}_{(1)} \, \Big|_{\text{crit.}} \, ,   \qquad\qquad X^\Lambda_{\text{NP}} = -\frac{\ii}{2} \,  X^{\Lambda}_{(2)} \, \Big|_{\text{crit.}} \, , \qquad \Lambda = 0 , \ldots, 3  \, . 
\ee 
Moreover, 
\be\label{eq:nontwistedentropycritical}
 S^{\text{KN-AdS$_{4}$}}_{\text{BH}} ( p^\Lambda , q_\Lambda , \cJ ) =  \cI_{\text{KN-AdS$_{4}$}} ( \fp^\Lambda , \bar \chi^\Lambda , \bar \omega) \, ,
\ee

\subsection{The purely electric STU model}

We first consider the case with $\fp = 0$, corresponding to the Kerr-Newman black holes in AdS$_4$ originally found in \cite{Cvetic:2005zi} for pairwise equal charges and generalized  to arbitrary charges in \cite{Hristov:2019mqp}.  We can set the magnetic charges to zero by choosing $\alpha = 0$ in \eqref{eq:C-parameterization}, as one can check from  \eqref{attr:eq:KNAdS4}. 

In the case with $\fp = 0$ we recover the entropy functional discussed in \cite{Choi:2018fdc}
\be
 \cE_{\text{\emph{e}KN-AdS$_{4}$}} ( \chi^\Lambda , \omega) = - 4 \ii \frac{\sqrt{\chi^0 \chi^1 \chi^2 \chi^3}}{\omega} = - \frac{2 \cF (\chi^\Lambda)}{\omega} \, ,
\ee
and we further retrieve
\be
 \big( X^\Lambda_{\text{SP}} \big)^* = - \frac{\ii}{2}\? \bar\chi^\Lambda \, , \qquad \quad
 X^\Lambda_{\text{NP}} = - \frac{\ii}{2}\? \bar\chi^\Lambda \, ,
\ee
at the critical point, which satisfies \eqref{eq:nontwistedentropycritical}.
This can be compared with \cite[(2.20)]{Choi:2018fdc} using the following dictionary,
\be
 \begin{aligned}
  & \omega_{\text{there}} = - \pi \omega_{\text{here}} \, , \qquad && \Delta_i = \ii \pi \chi^i \, , \quad &&& i = 1 , 2, 3 \, , \\
  & \Delta_4 = \ii \pi \chi^0 \, , && Q_4 = - \frac{1}{2 G_{\text{N}}} q_0 \, , \\
  & J = \frac{1}{2 G_{\text{N}}} \cJ \, , && Q_i = - \frac{1}{2 G_{\text{N}}} q_i \, , &&& i = 1 , 2, 3 \, , \\
  & g_{\text{there}} = \sqrt{2} \, , && g_{\text{here}} = 1 \, .
 \end{aligned}
\ee
Explicitly, in the notations of \cite{Choi:2018fdc} we have%
\footnote{It has been noticed in \cite{Cabo-Bizet:2018ehj,Cassani:2019mms} that there are two choices of constraints, $\sum_i \Delta_i - \omega = \pm 2 \pi \ii$, that lead to the same entropy and can be explained by the supersymmetry conditions for a class of Euclidean solutions. This observation extends to other cases and it is a consequence of the symmetries of the functional $\cI$. It would be interesting to see if there is a more physical explanation in terms of the attractor mechanism.}
\be\label{Choi}
 \cI_{\text{\emph{e}KN-AdS$_{4}$}} ( \chi^\Lambda , \omega) = \ii \frac{4 \sqrt{2} N^{3/2}}{3} \frac{\sqrt{\Delta_1 \Delta_2 \Delta_3 \Delta_4}}{\omega}
 + \sum_{i = 1}^{4} \Delta_i Q_i + \omega J + \lambda \bigg( \sum_{i=1}^4 \Delta_i - \omega - 2 \pi \ii \bigg) \, ,
\ee
where we used \eqref{AdS4:CFT3:dict} to translate $G_{\text{N}}^{(4)}$ into $N$.

A microscopic explanation for the entropy functional \eqref{Choi} was provided recently in \cite{Choi:2019zpz} by evaluating the three-dimensional superconformal index of ABJM and related theories.%
\footnote{See also \cite{Nian:2019pxj}.}
The method in \cite{Choi:2019zpz} does not use explicitly the holomorphic block picture \eqref{Zhol} but a closely related approach, which also uses the factorization
of the partition function. It indeed involves the gluing of two vortex generating functions, according to the same rules valid for holomorphic blocks. The field theory identity gluing uses
\bea
 X^{\Lambda}_{(1)} & = \chi^\Lambda  , \qquad && \omega_{(1)} = \omega \, , \\
 X^{\Lambda}_{(2)} & = - \chi^\Lambda \, , && \omega_{(2)} = -\omega  \, ,
\eea
and looks superficially different from the gravitational one \eqref{identitygluing}.
However, the Cardy and large $N$ limit of the vortex partition functions in \cite{Choi:2019zpz} are taken along particular directions in the complex plane of the chemical potentials and the final result is the same as \eqref{Choi}.\footnote{In particular, in the regime considered in \eqref{Choi} the minus sign in the chemical potential is equivalent to a complex conjugate. This might explain the complex conjugate we detect in gravity, see \eqref{eq:nontwistedsectionscritical}.} 
The main difficulty in comparing the gravitational and field theory identity gluing is that the known field theory computations use asymptotic expansions of special functions that are valid in a particular region in the complex plane.
Typically, the field theory identity gluing does not respect them and therefore further redefinitions of parameters are necessary.

\subsection[The dyonic \texorpdfstring{$X^0 X^1$}{X[0] X[1]} model]{The dyonic $X^0 X^1$ model}

The truncation is specified by the following identification of the sections and the charges
\bea
 & X^2 = X^0 \, , \qquad && X^3 = X^1 \, , \\
 & q_2 = q_0 \, , && q_3 = q_1 \, , \\
 & p^2 = p^0 \, , && p^3 = p^1 \, , \qquad p^1 = - p^0\, ,
\eea
where the last relation guarantees \eqref{P:constraint:KNAdS4}. From here on we set $p^0 \equiv - p$.
The attractor equations \eqref{attr:eq:KNAdS4} then read
\bea
 p & = - \frac{\alpha}{4 \alpha^2 - 4 \beta_0 \beta_1 + 1} \, , \\
 q_0 & = q_2 = \frac{\beta_0}{4 \alpha^2 - 4 \beta_0 \beta_1 + 1} \, , \\
 q_1 & = q_3 = \frac{\beta_1}{4 \alpha^2 - 4 \beta_0 \beta_1 + 1} \, ,
\eea
which can be easily solved for $(\alpha , \beta_1 , \beta_2)$.
The angular momentum is given by \eqref{JKNAdS4}, and can be easily rewritten as 

\be
 \label{JJ}
 \cJ = \frac{q_0 + q_1}{2} \left( 1 - \sqrt{1 - 16 \left( p^2 - q_0 q_1 \right)} \right) \, .
\ee
Notice that there are only three independent parameters in this model, and the angular momentum can be expressed in terms of the other charges.
Finally the entropy is given  by \eqref{SKNAdS4},%
\footnote{We correct a misprint in the formula for the entropy in \cite{Hristov:2019mqp}. Note also that we have redefined the angular momentum as $\cJ_{\text{there}} = - \frac{1}{2} \cJ_{\text{here}}$ (see footnote \ref{J:rscale:HKT}).}
\bea
 S_{\text{BH}}^{X^0 X^1} ( p , q_0 , q_1 , \cJ ) & = \frac{\pi}{4 G_{\text{N}}^{(4)} } \left( - 1 + \sqrt{1 - 16 \left( p^2 - q_0 q_1 \right)} \right) \\
 & = - \frac{\pi}{2 G_{\text{N}}^{(4)}} \frac{\cJ}{( q_0 + q_1 )} \, ,
\eea
where we take the branch of solutions considered  in \cite{Hristov:2019mqp}, \ie\;$q_{0, 1} > 0$. 

The horizon values of the sections at the SP and NP of the sphere can be computed from \eqref{XKNAdS4}. They read
\be
 \begin{aligned}
  X^0_{\text{SP}} & = - \ii \frac{\alpha ( 1 + 2 \alpha ) - ( \ii + 2 \beta_0 ) \beta_1}{1 - 4 \alpha^2 + 2 \ii \beta_1 + 2 \beta_0 ( \ii + 2 \beta_1 )} \, , \\
  X^1_{\text{SP}} & = \ii \frac{\alpha ( 1 - 2 \alpha ) +\beta_0 ( \ii + 2 \beta_1 )}{1 - 4 \alpha^2 + 2 \ii \beta_1 + 2 \beta_0 ( \ii + 2 \beta_1 )} \, , \\
  X^0_{\text{NP}} & = - \ii \frac{\alpha ( 1 - 2 \alpha ) - ( \ii - 2 \beta_0 ) \beta_1}{1 - 4 \alpha^2 - 2 \ii \beta_1 - 2 \beta_0 ( \ii - 2 \beta_1 )} \, , \\
  X^1_{\text{NP}} & = \ii \frac{\alpha ( 1 + 2 \alpha ) + \beta_0 ( \ii - 2 \beta_1 )}{1 - 4 \alpha^2 - 2 \ii \beta_1 - 2 \beta_0 ( \ii - 2 \beta_1 )} \, .
 \end{aligned}
\ee
Our $\cE$-functional reduces to that of \cite[Eq.\,(1)]{Hristov:2019mqp} in this case, \ie\;
\be
 \cE_{\text{KN-AdS$_{4}$}} ( \fp , \chi^\Lambda , \omega) \Big|_{X^0 X^1}= \frac{4 \ii \chi^0 \chi^1}{\omega} + \ii \omega \fp^2 \, ,
\ee
where we chose to work with the negative determination for the square roots in \eqref{eq:entropyfunctionalnotwist2}, which correctly reproduces the branch of solutions we are looking at.
Recall that, $\fp^0 = \fp^2 = - \fp^1= - \fp^3 = - \fp$ and $\fp=-2 p$.
We thus need to extremize
\be
 \label{KNAdS4:X0X1:I-functional}
 \cI_{X^0 X^1} ( \fp , \chi^\Lambda , \omega) 
 =  \frac{\pi}{4 G_{\text{N}}^{(4)}} \bigg( \frac{4 \ii \chi^0 \chi^1}{\omega} + \ii \omega \fp^2 - 4 \ii \sum_{i=0}^1 q_i \chi^i - 2 \omega \cJ \bigg) \, ,
\ee
under the constraint
\be
 \chi^0 + \chi^1 - \frac{\ii}{2} \omega = 1 \, .
\ee
The extremization equations read
\bea
  \frac{\partial \cI_{X^0 X^1}}{\partial \chi^0} & = 0 = \left( 2 ( q_0 - q_1 ) - \ii \right) \omega + 4 \chi^0 - 2 \, , \\
  \frac{\partial \cI_{X^0 X^1}}{\partial \omega} & = 0 = 4 ( \chi^0 - 1 ) \chi^0 + \omega^2 \left( \fp^2 + 2 \ii ( \cJ - q_1 ) \right)  .
\eea
The critical points are then given by
\bea
 \bar\chi^0 & = \frac{1}{2} \left( 1 \mp \frac{\ii - 2 ( q_0 - q_1 )}{\sqrt{4 \left( 2 \ii \cJ + \fp^2 \right) + 4 \left( ( q_0 - q_1 )^2 - \ii ( q_0 + q_1 ) \right) - 1}} \right) , \\
 \bar\omega & = \mp \frac{2}{\sqrt{4 \left( 2 \ii \cJ + \fp^2 \right) + 4 \left( ( q_0 - q_1 )^2 - \ii ( q_0 + q_1 ) \right) - 1}} \, ,
\eea
and the value of the entropy functional \eqref{KNAdS4:X0X1:I-functional} at its critical point is found to be
\bea
 \cI_{X^0 X^1} \Big|_{\text{crit.}} = \frac{\pi}{4 G_{\text{N}}^{(4)}} \left( - 1 - 2 \ii ( q_0 + q_1 ) \mp \ii \sqrt{4 \left( 2 \ii \cJ + \fp^2 \right) + 4 \left( ( q_0 - q_1 )^2 - \ii ( q_0 + q_1 ) \right) - 1} \right) .
\eea
Although not immediately obvious from this expression, upon using the constraint \eqref{JJ} and taking the solution that leads to a real positive entropy, one obtains
\be
 S_{\text{BH}}^{X^0 X^1} ( p , q_0 , q_1 , \cJ ) =  \cI_{X^0 X^1} ( \fp , \bar\chi^0 , \bar\omega) \, .
\ee

\subsection{The general case}

We can also consider the entropy functional \eqref{eq:entropyfunctionalnotwist} in the general case of taking the symplectic vector $\cC$ in \eqref{eq:C-parameterization} without further simplifications. The resulting formulae for the conserved charges $\Gamma$ in terms of the parameters $(\alpha, \beta_\Lambda)$, as well as subsequent expressions for the sections and entropy, are not really presentable and offer no specific insight. We have checked numerically that the expected relations \eqref{eq:nontwistedsectionscritical} and \eqref{eq:nontwistedentropycritical} hold once again, for a suitable  choice of determination of the square root in \eqref{eq:entropyfunctionalnotwist2}, letting us to conclude the proposed attractor mechanism is valid in full generality for these solutions.

A field theory explanation for the entropy functional \eqref{eq:entropyfunctionalnotwist} for generic magnetic charges is still missing.
It would be interesting to provide it using \eqref{Zhol} or the factorization method of \cite{Choi:2018fdc}.

\section[Kerr-Newman-AdS\texorpdfstring{$_5$}{(5)}]{Kerr-Newman-AdS$_{5}$}
\label{sec:KNAdS5}

Our last example deals with the Kerr-Newman black holes in AdS$_5$.
The five-dimensional solutions in minimal gauged supergravity were first found in \cite{Gutowski:2004ez} and generalized to two rotations in \cite{Chong:2005hr}.
The most general solutions of the STU model were then spelled out in \cite{Kunduri:2006ek}.
The near horizon geometry is a fibration of AdS$_2$ over a non-homogeneously squashed three-sphere \cite{Kunduri:2007qy}.
The holographically dual four-dimensional boundary theory remains superconformal in this case.
Various results  have been obtained recently in evaluating the superconformal index of the dual field theory on $S^3 \times S^1$ \cite{Kinney:2005ej,Romelsberger:2005eg}
in various limits and matching the answer to the macroscopic entropy \cite{Cabo-Bizet:2018ehj,Choi:2018hmj,Benini:2018ywd}.%
\footnote{See \cite{,Honda:2019cio,ArabiArdehali:2019tdm,Kim:2019yrz,Cabo-Bizet:2019osg,Amariti:2019mgp,Larsen:2019oll,Lezcano:2019pae,Lanir:2019abx} for further developments.}

For our present purposes we shall consider the 4D/5D connection, as done in \cite{Hosseini:2017mds}.
In this case in order to preserve supersymmetry one is led to do a more general Scherk-Schwarz reduction \cite{Andrianopoli:2004im,Looyestijn:2010pb}
and supersymmetry in the lower-dimensional theory is thus preserved with a partial topological $A$-twist on the $S^2$ inside the original $S^3$.
From a four-dimensional perspective the KN-AdS$_5$ black holes therefore fit in the class of solutions of \cite{Hristov:2018spe} of rotating attractors with a twist.

We consider the electric STU model with prepotential
\be
 \cF (X^\Lambda) = \frac{X^1 X^2 X^3}{X^0} \, ,
\ee
and purely electric gauging coming from the Scherk-Schwarz reduction \cite{Hosseini:2017mds,Hristov:2018spe}
\be
 G = \{g^{\Lambda}; \sqrt{2} \cosh (\delta) , g_i \} \, , \qquad g_i \equiv g \, , \qquad g^\Lambda = 0 \, .
\ee
We set $g=\sqrt{2}$.

The dimensional reduction of the black hole in \cite{Kunduri:2006ek}  to four dimensions was already performed in \cite{Hristov:2018spe}.  The set of four-dimensional electromagnetic charges 
\be
 \Gamma = \{ p^\Lambda ; q_\Lambda \} \, ,
\ee
can be written in terms of  the variables $\mu_i$ and $\Xi_{a,b}$ appearing in the original solution \cite{Kunduri:2006ek}  as \cite{Hristov:2018spe} 
\be
 p^0  = -\frac{1}{\sqrt{2}\? \cosh (\delta)} \, , \qquad
 p^i  = 0 \, , \quad i = 1 ,2 , 3 \, ,
\ee
and
\bea
 q_0 &= -\frac{1}{4 \sqrt{2} \cosh(\delta)} \left( \Big( 1 + \gamma_1+ \frac{\gamma_2}{2} \Big) \cosh (\delta) - (1+\gamma_1+\gamma_2+\gamma_3) \right) \, , \\
 q_i &= -\frac{1}{4 \sqrt{2} \cosh(\delta)} \left( \mu_i + \frac{\gamma_2}{2} - \frac{\gamma_3}{\mu_i} \right) \, , \qquad i = 1 , 2 , 3 \, .
\eea
For a lighter notation we defined
\be
 \gamma_1 \equiv \mu_1 + \mu_2 + \mu_3 \, , \qquad  \gamma_2 \equiv \mu_1 \mu_2 + \mu_1 \mu_3 + \mu_2 \mu_3 \, , \qquad \gamma_3 \equiv \mu_1 \mu_2 \mu_3 \, ,
\ee
and
\be
 \sinh(\delta) = \frac{\Xi_a - \Xi_b}{2 \sqrt{\Xi_a \Xi_b}} \, , \qquad \cosh (\delta) = \frac{\Xi_a + \Xi_b}{2 \sqrt{\Xi_a \Xi_b}} \, .
\ee
Notice that  the angle along which we reduce in the solution of \cite{Kunduri:2006ek} has period $4\pi \cosh (\delta)$.
As already mentioned, the reduction along the Hopf fibre of $S^3$ introduces a magnetic charge $p^0$. The theory is thus topologically twisted with the twisting condition
\be
 g_0 p^0 =-1 \, .
\ee

The quartic invariant is again given by \eqref{eq:I4electricstu} and the main features of the solution are summarized in appendix \ref{app:A2}. The four-dimensional near horizon solution for the parameters ${\rm j}, \cH_0$ is spelled out in \cite[sect.\,4.3.2]{Hristov:2018spe}. In terms of the variables $\delta$ and $\mu_i$, the Bekenstein-Hawking entropy is given by
\be
 S^{\text{KN-AdS$_{5}$}}_{\text{BH}} \left( p^0 , q_i , \cJ \right) = \frac{\pi^2}{4 G^{(5)}_{\text{N}}} \Pi(\delta, \mu_i)  \, ,
\ee
where we defined
\be
 \Pi \equiv 2 \sqrt{ \gamma_3 ( 1 + \gamma_1) - \frac14 \gamma_2^2 + 2 ( 1 - \cosh(\delta) ) \prod_{i=1}^3 ( 1 + \mu_i )} \, ,
\ee
and
\be
 \frac{1}{G^{(4)}_{\text{N}}} = \frac{4 \pi \cosh (\delta)}{G^{(5)}_{\text{N}}} \, .
\ee

The angular momentum in four dimensions is given by
\be
 \cJ = - \frac{\sinh (\delta)}{8 \cosh^2 (\delta)} \left( 1+\gamma_1+\frac{\gamma_2}{2} \right) ,
\ee
and its corresponding chemical potential reads
\be
 w = \frac{\sinh ( 2 \delta )}{\Pi(\delta, \mu_i)} \, .
\ee
We can also write down the symplectic sections at the near horizon region if we further use the definition
\be
 \Theta(\delta, \mu_i) \equiv 3 - 2 \cosh (\delta) + \gamma_1 \, .
\ee
At the NP and SP of the sphere ($\theta = \pi$ and $\theta = 0$, respectively), we find
\bea
 X^0_{\text{SP}} - X^0_{\text{NP}} &= - 2 w p^0 \, ,\\
 X^0_{\text{SP}} + X^0_{\text{NP}} & = \frac{2 \sqrt{2}}{\Theta} \left( \frac{\ii}{2} - \frac{1 + \gamma_1 + \frac{\gamma_2}{2} - \Theta \cosh ( \delta )}{\Pi } \right) .
\eea
We also obtain 
\bea
 X^i_{\text{SP}} = \frac{1}{\sqrt{2} \Theta \Pi } & \bigg( \ii \Pi (\Theta - 1 - \mu_i) + \frac{\gamma_3}{\mu_i} ( 2 + \gamma_1 - \mu_i ) + \mu_i \bigg( \mu_{i}^2 - \sum_{i = 1}^3 \mu_{i}^2 \bigg) \\
 & - 4 \left( 1 - \cosh ( \delta ) \right) \left( 1 + \mu_i + \frac{\gamma_2}{2} - \frac{\gamma_3}{\mu_i} \right) \! \bigg) \, , \qquad i = 1 , 2 , 3 \, ,
\eea
and
\be
 X^i_{\text{SP}} - X^i_{\text{NP}} = - 2 w p^i = 0 \, , \qquad i = 1 , 2 , 3 \, .
\ee
Remarkably, they satisfy the constraint
\be
 g_\Lambda \left( X^\Lambda_{\text{SP}} + X^\Lambda_{\text{NP}} \right) + 2 w \tanh ( \delta ) = 2 \ii \, .
\ee

\subsection{Attractor mechanism}

Since from the four-dimensional point of view the theory is topologically twisted, we need to use the $A$-gluing \eqref{Agluing}:
\bea
 X^{\Lambda}_{(1)} & = \chi^\Lambda + \ii \frac{\omega}{2} \fp^\Lambda \, , \qquad && \omega_{(1)} = \omega \, , \\
 X^{\Lambda}_{(2)} & = \chi^\Lambda - \ii \frac{\omega}{2} \fp^\Lambda \, , && \omega_{(2)} = - \omega \, ,
\eea
where, to keep the same notations as before, we define $\fp^\Lambda = - 2 p^\Lambda$. This gives the $\cE$-functional 
\be
 \label{KN-AdS5:cE:4D}
 \cE_{\text{KN-AdS$_{5}$}} ( \fp^0 , \chi^\Lambda , \omega) = 4 \ii \frac{\fp^0 \chi^1 \chi^2 \chi^3}{(2 \chi^0 )^2 + ( \omega \fp^0 )^2} \, .
\ee
The entropy functional \eqref{gluing}  is  given by 
\bea
 \label{I-functional:GR}
 \cI_{\text{KN-AdS$_{5}$}} ( \fp^0 , \chi^\Lambda , \omega ) & = \frac{\pi}{4 G_{\text{N}}^{(4)}}\left (\cE_{\text{KN-AdS$_{5}$}} ( \fp^0 , \chi^\Lambda , \omega)
 - 2 \ii \chi^\Lambda q_\Lambda - 2 \omega \cJ\right ) \\
 & + \lambda \left( g_\Lambda \chi^\Lambda - 2 - \ii \omega \tanh ( \delta ) \right) \, .
\eea
Evaluating the $\cI$-functional \eqref{I-functional:GR} at its critical point we recover
\be
 S^{\text{KN-AdS$_{5}$}}_{\text{BH}} \left( p^0 , q_i , \cJ \right) = \cI_{\text{KN-AdS$_{5}$}} ( \fp^0 , \bar \chi^\Lambda , \bar \omega )
\, ,
\ee
and the critical values $(\bar\chi^\Lambda , \bar\omega)$ read
\bea
 \bar{\chi}^\Lambda & = - \ii \left( X^\Lambda_{\text{SP}} + X^\Lambda_{\text{NP}} \right) \, , \qquad \Lambda = 0, \ldots, 3 \, , \\
 \bar{\omega} & = - \sqrt{2} \cosh (\delta ) \left( X^0_{\text{SP}} - X^0_{\text{NP}} \right) = - 2 w \, .
\eea
The values of the sections at the SP and the NP are then given by the usual attractor relations
\be
 \begin{aligned}
  & X^0_{\text{SP, NP}} = \frac{\ii}{2} \left( \bar \chi^0 \pm \ii \frac{\bar \omega}{2} \fp^0 \right) \, , \\
  & X^i_{\text{SP}} = X^i_{\text{NP}} = \frac{\ii}{2} \bar{\chi}^i \, ,  \qquad i = 1 , 2, 3 \, .
 \end{aligned}
\ee

Notice that we have a dependence on the parameter $\delta$, which is related to the conserved charges in five dimensions, in the gaugings and in the constraint that the chemical potentials fulfill.
This is an artifact of the dimensional reduction, in order to obtain a consistent supersymmetric four-dimensional solution.
We nevertheless see that the gluing procedure nicely works at the formal level.

\subsection{Comparison with the five-dimensional entropy functional} \label{sec:5dpicture}

The entropy functional \eqref{I-functional:GR}  is equivalent to the functional for KN-AdS$_5$ black holes found by \cite{Hosseini:2017mds} in a different basis.
In \eqref{I-functional:GR} the Legendre transform is done with respect to a four-dimensional basis of charges.
The latter and the corresponding chemical potentials are related to the natural five-dimensional ones by a linear redefinition involving $\delta$.

The entropy functional found in \cite{Hosseini:2017mds} is given in \eqref{HHZ}. Reinstating the five-dimensional Newton's constant, we can write it as
\be
 \label{HHZ2}
 \cI(\Delta_a,\omega_i)=  \ii \frac{\pi^2}{2G^{(5)}_{\text{N}}} \? \frac{\Delta^1\Delta^2\Delta^3}{\omega_1\omega_2} + 2 \pi \ii \left (\sum_{a=1}^3 \Delta^a Q_a -  \sum_{i=1}^2 \omega_i J_i\right) ,
\ee
where we used the holographic dictionary
\be
 \frac{1}{G^{(5)}_{\text{N}}} = \frac{2}{\pi} N^{2} \, .
\ee
 
The five-dimensional charges used in \cite{Hosseini:2017mds} are related to the four-dimensional conserved charges by%
\footnote{In order to compare with \cite[(4.18)]{Hosseini:2017mds} one has to set $\delta = 0$, $\cJ = 0$ here, that corresponds to KN-AdS$_5$ black holes with equal angular momenta $(J_1 = J_2)$, and therefore a static black hole in four dimensions. Moreover, $g_{\text{here}} = \sqrt{2}$ while $g_{\text{there}}$ = 1. This leads to the following redefinition of four-dimensional charges: $q_\Lambda^{\text{there}} = \sqrt{2} \? q_\Lambda^{\text{here}}$, $\Lambda = 0, \ldots, 3$.}
\be
 \begin{aligned}
  Q_i^{(5)} & = - \frac{\sqrt{2}\? \pi \cosh (\delta)}{G^{(5)}_{\text{N}}}\? q_i \, , \qquad i = 1 , 2 , 3 \, , \\
   J_{1,2} & = \frac{\sqrt{2} \? \pi e^{\pm \delta } \cosh ( \delta )}{G^{(5)}_{\text{N}}} \left( q_0 \mp \sqrt{2} \cosh ( \delta ) \cJ \right) .
%
%
 \end{aligned}
\ee
We see that the entropy functional  \eqref{I-functional:GR} matches precisely with the one presented in \cite{Hosseini:2017mds}, upon the identification
\bea
 \Delta^i & = \frac{1}{\sqrt{2}} \chi^i \, , \qquad i = 1 , 2 , 3 \, , \\
 \omega_{1,2} & = \frac{1}{4} \left( 1 \mp \tanh ( \delta ) \right) \left( \sqrt{2} \cosh ( \delta ) \chi^0 \pm \ii \omega \right) .
\eea

\section{Generalization to higher dimensions}
\label{sec:higher}

We expect that our construction can be extended to other dimensions. We give few examples here leaving a detailed analysis for a later publication \cite{Hosseini:2019HHZ2}.
The  gravitational blocks in higher dimensions are of the form
\bea\label{Bhigherdim}
 & \cB_{\text{5D}} (X^\Lambda , \omega_i) \equiv - \frac{\cF_{\text{5D}}(X^\Lambda)}{\omega_1 \omega_2} \, , \\
 & \cB_{\text{6D}} (X^\Lambda , \omega_i) \equiv - \frac{\cF_{\text{6D}}(X^\Lambda)}{\omega_1 \omega_2} \, , \\
 & \cB_{\text{7D}} (X^\Lambda , \omega_i) \equiv - \frac{\cF_{\text{7D}}(X^\Lambda)}{\omega_1 \omega_2 \omega_3} \, ,
\eea
where $\omega_i$ are chemical potentials associated with rotations. We can also see them as equivariant parameters associated to the independent rotational symmetries
of the solutions, which we assume to be the maximal ones in each dimension (two, two and three, respectively). 

\subsection{Five dimensions}

Let us first consider the five-dimensional picture.
For KN-AdS$_5$ black holes with near horizon region AdS$_2 \times_w S^3$, and rotating black strings in AdS$_5$ with near horizon BTZ$ \times_w S^2$ we use
\be
 \cF_{\text{5D}} (X^\Lambda) = X^1 X^2 X^3 \, . 
\ee
From the gravitational point of view we glue two copies of $D_2\times S^1$, where $D_2$ is a disk. It is easy to see that, with suitable redefinition of variables, the entropy functional  \eqref{HHZ} for KN-AdS$_5$ can be obtained by gluing the two copies into an $S^3$ with the identifications
\bea
 \label{5did}
 & X^{a}_{(1)} = \chi^a \, , \qquad && \omega_{1,(1)} = \omega_1 \, , \qquad &&& \omega_{2,(1)} = \omega_2 \, ,\\
 & X^{a}_{(2)} = \chi^a \, , && \omega_{1,(2)} = \omega_1  \, , &&& \omega_{2,(2)} = \omega_2 \, ,
\eea
while the entropy functional  \eqref{BSI} for rotating black strings an be obtained by gluing the two copies into an $S^2\times S^1$, where the $S^1$ lies inside BTZ, with the identifications
\bea
 \label{5dA}
 & X^{\Lambda}_{(1)} = \chi^\Lambda + \ii \frac{\omega_1}{2} \? \fp^\Lambda \, , \qquad && \omega_{1, (1)} = \omega_1 \, , \qquad &&& \omega_{2, (1)} = \omega_2 \, , \\
 & X^{\Lambda}_{(2)} = \chi^\Lambda - \ii \frac{\omega_1}{2} \? \fp^\Lambda \, , && \omega_{1, (2)} = - \omega_1 \, , \qquad &&& \omega_{2, (2)} = \omega_2\, .
\eea
The partition functions of the holographically dual field theories are obtained by gluing copies of the $D_2 \times T^2$ partition function, the four-dimensional holomorphic blocks. The latter have the expansion
\be
 \label{Bethehol2}
 B^\alpha ( \Delta^\Lambda | \omega_1 ) \underset{\omega_1 \to 0}{\sim} \exp \bigg( \! -\frac{1}{\omega_1} \cW ( x^\alpha , \Delta^\Lambda ) \bigg) \, ,
\ee
where $\omega_1$ is the equivariant parameter on $D_2$. Moreover, in a Cardy limit associated with the torus $T^2$, the twisted superpotential of $\cN = 4$ SYM  reads \cite[(3.23)]{Hosseini:2016cyf} 
\be
 \wt \cW ( \Delta^\Lambda | \beta) = \ii \pi^3 N^2 \? \frac{\Delta^1 \Delta^2 \Delta^3}{2 \beta} \, ,
\ee
with $\sum_{\Lambda=1}^3 \Delta^\Lambda = 2$. Here $\beta = - 2 \pi \ii \tau$ where $\tau$ is the modulus of the torus.\footnote{The twisted superpotential is evaluated on the Bethe vacuum that dominates the saddle point approximations
of both the topologically twisted index \cite{Hosseini:2016cyf} and the superconformal ones \cite{Benini:2018ywd}.}
By identifying $\Delta^\Lambda$ with $X^\Lambda$ and $\beta$ with $\omega_2$,
 it is not hard to recognize in the exponent of \eqref{Bethehol2} the expression of the five-dimensional gravitational block $\cB_{5\text{D}}$.
The gluing \eqref{5dA} precisely corresponds to the field theory gluing used for the topologically twisted index in \cite{Nieri:2015yia}.
As in four dimensions, the gluing \eqref{5did} is superficially different from the $S$-gluing used in field theory \cite{Nieri:2015yia},%
\footnote{See also \cite{Yoshida:2014qwa,Peelaers:2014ima}.}
but we again expect to have the same physical effect in the saddle point approximation. It would be interesting to relate the two more directly.

\subsection{Six dimensions}

Let us now consider the six-dimensional point of view.  We consider here black holes in the AdS$_6 \times_w S^4$ background of massive type IIA supergravity \cite{Brandhuber:1999np}. 
For KN-AdS$_6$ \cite{Chow:2008ip}, \emph{m}AdS$_6$ \cite{Hosseini:2018usu,Suh:2018szn} and possible rotating generalizations with near horizon AdS$_2 \times_w \cM_4$,
where the manifold $\cM_4$ is either $S^4$ or toric K\"ahler, we should use\footnote{The variables $X_1$ and $X_2$ can be associated with the two isometries of the internal manifold $S^4$ of the solution \cite{Brandhuber:1999np}.}
\be
 \label{F6d}
 \cF_{\text{6D}} (X^\Lambda) = (X^1 X^2)^{3/2} \, . 
\ee
The structure of five-dimensional supersymmetric partition functions and their decomposition in terms of holomorphic blocks are not fully understood.
It was argued in \cite{Hosseini:2018uzp,Crichigno:2018adf} that the effective Seiberg-Witten prepotential should play the role of the twisted superpotential for three- and four-dimensional field theories.
The expression \eqref{F6d} is precisely the critical value of the Seiberg-Witten prepotential of the dual field theory in the large $N$ limit, see \cite[(3.71)]{Hosseini:2018uzp}.
A natural conjecture inspired by \cite{Nekrasov:2003vi,Bawane:2014uka,Bershtein:2015xfa,Hosseini:2018uzp} is that we need a gluing of the form
\be
 \label{conjecture:gluing}
 \cE ( \chi^\Lambda , \omega_i ) = \sum_{\sigma = 1}^{\chi_{\text{E}} (\cM_4)} \cB_{\text{6D}}  \big(X^\Lambda_{(\sigma)} , \omega_{i, (\sigma)} \big) \, ,
\ee
where $\chi_{\text{E}} (\cM_4)$ denotes the Euler characteristic of $\cM_4$.
The contributions are associated to the NP and SP of $S^4$ and to the fixed points under the torus action for a toric manifold $\cM_4$ in the spirit of \cite{Nekrasov:2003vi}.

The entropy functional for Kerr-Newman black holes in AdS$_6$ was found in \cite[(3.15)]{Choi:2018fdc} and it is given by
\be
 \label{KNAdS6:EF}
 \cI_{\text{KN-AdS$_6$}} (\Delta , \omega_i) = - \ii \frac{\pi}{(3 g)^4 G_{\text{N}}^{(6)}} \frac{\Delta^3}{\omega_1 \omega_2} + \Delta Q + \sum_{i=1}^2 \omega_i J_i + \lambda \bigg( \Delta - \sum_{i=1}^2 \omega_i - 2 \pi \ii \bigg) \, ,
\ee
where $Q$ is the electric charge and $J_i$, $i=1,2$, are the two angular momenta of the solution.
$\Delta$ and $\omega_i$ are, respectively, the chemical potentials conjugate to these conserved charges.
Notice that only one of the two possible electric charges is turned on in the solution  \cite{Chow:2008ip}.
The entropy functional \eqref{KNAdS6:EF} fits in our general formalism by simply identity gluing the two gravitational blocks $\cB_{\text{6D}}$ as follows:
\bea
 & X^\Lambda_{(1)} = \chi^\Lambda \, , \qquad && \omega_{1, (1)} = \omega_1 \, , \qquad &&& \omega_{2, (1)} = \omega_2 \, , \\
 & X^\Lambda_{(2)} = \chi^\Lambda \, , && \omega_{1, (2)} = \omega_1 \, , &&& \omega_{2, (2)} = \omega_2 \, .
\eea
This leads to
\be
 \cE_{\text{KN-AdS$_6$}} (\chi^\Lambda , \omega_i) = -\frac{2 ( \chi^1 \chi^2 )^{3/2}}{\omega_1 \omega_2} \, ,
\ee
that, up to a normalization, can be clearly mapped to \eqref{KNAdS6:EF} upon identifying $\chi^1= \chi^2 \equiv \Delta$.

Another interesting example is the class of static \emph{m}AdS$_6$ black holes found in \cite{Hosseini:2018usu,Suh:2018szn}. 
The entropy functional of this class of black holes when the near horizon geometry is AdS$_2 \times \cM_4$, with $\cM_4$ being a K\"ahler-Einstein manifold, reads \cite[(6.8)]{Hosseini:2018usu}
\be
 \label{I-functional:smAdS6}
 \cI_{\text{\emph{sm}AdS$_6$}} (\fp^I , \Delta^I) = \frac{\Vol ( \cM_4 )}{( 3 \sqrt{2} )^4 G^{(6)}_{\text{N}}} \sum_{I , J = 1}^2 \fp^I \fp^J \frac{\partial^2 (\Delta^1 \Delta^2)^{3/2}}{\partial \Delta^I \partial \Delta^J} - \lambda \bigg( \sum_{I = 1}^2 \Delta^I - 2 \bigg) \, ,
\ee
where $\fp^1+\fp^2=2\kappa$ if the metric is normalized as $R_{\mu\nu}=\kappa g_{\mu\nu}$.
We now show that we can reproduce the above entropy functional  by gluing  six-dimensional gravitational blocks.
Unfortunately, there are no regular black hole solutions with manifolds $\cM_4$ of positive curvature \cite{Hosseini:2018usu}, as one can see by extremizing \eqref{I-functional:smAdS6}.
Nevertheless,  it makes sense to  consider all kind of  horizons  because we want to reproduce the {\it functional form} of $ \cI_{\text{\emph{sm}AdS$_6$}} (\fp^I , \Delta^I)$, independently of whether it has acceptable critical points or not.

Let us then focus on the case where $\cM_4$ is the complex projective space $\bP^2$, that is a toric manifold also.
Denote the  generators of the $(\bC^*)^2$ action on the tangent space at the three fixed points $P_{(l)}$  by $\omega_{1,(l)}$, $\omega_{2,(l)}$ with $l = 1 , 2 , 3$.
Since $\chi_{\text{E}} (\bP^2) = 3$ we should fuse three copies of $\cB_{\text{6D}}$ into each other using the higher-dimensional $A$-gluing as follows (see \cite[Example.\,2.1]{Hosseini:2018uzp}):
\be
 X^\Lambda_{(l)} = \chi^\Lambda + \ii \frac{\omega_{1,{(l)}}}{2} \? \fp^\Lambda + \ii \frac{\omega_{2,{(l)}}}{2} \? \fp^\Lambda \, , \qquad l = 1, 2 , 3 \, ,
\ee
with
\bea
 & \omega_{1, (1)} = \omega_1 \, , \qquad \qquad \, \? \? \omega_{2, (1)} = \omega_2 \, , \\
 & \omega_{1, (2)} = \omega_2 - \omega_1 \, , \qquad \? \omega_{2, (2)} = - \omega_1 \, , \\
 & \omega_{1, (3)} = - \omega_2 \, , \qquad \qquad \! \!  \omega_{2, (3)} = \omega_1 - \omega_2 \, .
\eea
Thus, our $\cE$-functional \eqref{conjecture:gluing} for $\cM_4 = \bP^2$ reads
\be
 \cE ( \fp^I , \chi^I ) = \frac{9}{8} \sum_{I , J = 1}^2 \fp^I \fp^J \frac{\partial^2 (\chi^1 \chi^2)^{3/2}}{\partial \chi^I \partial \chi^J} \, ,
\ee
which is, up to a normalization, \eqref{I-functional:smAdS6} upon identifying $\chi^I$ with $\Delta^I$.

We can also consider the case of $\cM_4 = S^2 \times S^2$.%
\footnote{Also in this case, no static \emph{m}AdS$_6$ black hole exists with this horizon topology. There are solutions with horizon $\Sigma_{\fg_1}\times \Sigma_{\fg_2}$, where $\Sigma_\fg$ denotes a Riemann surface of genus $\fg$, whenever $\fg_1 >1$ or $\fg_2>1$.} 
The entropy functional reads \cite[(5.10)]{Hosseini:2018usu}
\be
 \label{TTI:S2xS2}
 \cI_{S^2 \times S^2 \times S^1} (\fs^I , \ft^I , \Delta^I) =
 \frac{(2 \pi)^2}{81 G^{(6)}_{\text{N}}}  \sum_{I , J = 1}^2 \fs^I \ft^J \frac{\partial^2 (\Delta^1 \Delta^2)^{3/2}}{\partial \Delta^I \partial \Delta^J}
 - \lambda \bigg( \sum_{I=1}^2 \Delta^I - 2 \bigg) \, ,
\ee
where $\fs^I$, $\ft^I$ are the magnetic charges on the two $S^2$ and they satisfy the quantization conditions $\fs^1 + \fs^2 = 2$ and $\ft^1 + \ft^2 = 2$.
This result has been also derive from field theory using  the topologically twisted index of the dual five-dimensional $\cN=1$ theory 
on $S^2 \times S^2 \times S^1$ in \cite{Hosseini:2018uzp}.%
\footnote{This is the $\USp(2N)$ gauge theory with $N_f$ fundamental flavors and an antisymmetric matter field, which has a five-dimensional UV fixed point with enhanced $E_{N_f+1}$ global symmetry \cite{Seiberg:1996bd}. The holographic dictionary reads \cite{Jafferis:2012iv}, $G_N^{(6)} = \frac{5 \pi}{27 \sqrt{2}} \frac{\sqrt{8 - N_f}}{N^{5/2}}$.} 
The above functional can be easily obtained by gluing four copies, since $\chi_{\text{E}}(S^2 \times S^2) = 4$, of six-dimensional gravitational blocks as follows:
\bea
& X^\Lambda_{(1)} = \chi^\Lambda + \ii \frac{\omega_{1, (1)}}{2} \fs^\Lambda + \ii \frac{\omega_{2,(1)}}{2} \ft^\Lambda \, , \qquad
& X^\Lambda_{(2)} = \chi^\Lambda + \ii \frac{\omega_{1, (2)}}{2} \ft^\Lambda + \ii \frac{\omega_{2,(2)}}{2} \fs^\Lambda \, , \\
& X^\Lambda_{(3)} = \chi^\Lambda + \ii \frac{\omega_{1, (3)}}{2} \fs^\Lambda + \ii \frac{\omega_{2,(3)}}{2} \ft^\Lambda \, , \qquad
& X^\Lambda_{(4)} = \chi^\Lambda + \ii \frac{\omega_{1, (4)}}{2} \ft^\Lambda + \ii \frac{\omega_{2,(4)}}{2} \fs^\Lambda \, ,
\eea
where (see \cite[Example.\,2.2]{Hosseini:2018uzp})
\be
 \begin{aligned}
  & \omega_{1,(1)} = \omega_1 \, , \qquad ~~\; \? \omega_{2,(1)} = \omega_2 \, , \\
  & \omega_{1,(2)} = \omega_2 \, , \qquad ~~\; \? \omega_{2,(2)} = - \omega_1 \, , \\
  & \omega_{1,(3)} = -\omega_1 \, , \qquad \, \? \omega_{2,(3)} = -\omega_2 \, , \\
  & \omega_{1,(4)} = -\omega_2 \, , \qquad \, \? \omega_{2,(4)} = \omega_1 \, .
 \end{aligned}
\ee
The four contributions correspond to the four fixed points of the torus action associated with the poles of the spheres.
Thus, our $\cE$-functional \eqref{conjecture:gluing} for $\cM_4 = S^2 \times S^2$ reduces to
\be
 \cE ( \fs^I , \ft^I , \chi^I ) = \sum_{I , J = 1}^2 \fs^I \ft^J \frac{\partial^2 (\chi^1 \chi^2)^{3/2}}{\partial \chi^I \partial \chi^J} \, ,
\ee
which is, up to a normalization, \eqref{TTI:S2xS2} upon identifying $\Delta^I$ with $\chi^I$.

We expect the existence of other static and rotating six-dimensional black holes with two isometries and positive real entropy. 
Our discussion leads to a prediction for the entropy of these objects.

\subsection{Seven dimensions}

Let us finally consider the seven-dimensional perspective.
For KN-AdS$_7$ \cite{Cvetic:2005zi,Chow:2007ts} with near horizon AdS$_2 \times_w S^5$, AdS$_7$ black strings \cite{Benini:2013cda} and possible rotating generalizations with near horizon BTZ$ \times_w \cM_4$ we should use (\cf\;\cite[(1.4)]{Hosseini:2018dob} and \cite[(3.22)]{Hosseini:2018uzp})
\be\label{7D}
 \cF_{\text{7D}} (X^\Lambda) = (X^1 X^2)^{2} \, . 
\ee
It is easy to see that the entropy functionals found  in  \cite{Hosseini:2018dob} and \cite{Hosseini:2018uzp} for KN-AdS$_7$ and AdS$_7$ black strings, respectively, can be obtained by gluing blocks of this form.

In principle, rotating black holes in \emph{m}AdS$_6$, KN-AdS$_6$ black holes, rotating AdS$_7$ black strings,
and KN-AdS$_7$ black holes can be all studied in F(4) gauged supergravity coupled to vector multiplets \cite{DAuria:2000xty,Andrianopoli:2001rs} using a six-dimensional point of view.
For example, after Scherk-Schwarz reduction along the Hopf fiber of $S^5$ the near horizon of KN-AdS$_7$ becomes AdS$_2 \times_w \bP^2$
and the six-dimensional black hole becomes topologically twisted. We would then expect to recover the entropy functional for KN-AdS$_7$ by gluing
three six-dimensional gravitational blocks $\cB_{6\text{D}}$ associated to the fixed points of the toric action on $\bP^2$ according to \eqref{conjecture:gluing}.
It would be interesting to provide a unifying description of all six- and seven-dimensional black objects using six-dimensional supergravity.
This would be in the spirit of the analysis that we have performed in this paper for four- and five-dimensional black objects.  

We hope to have given a glimpse of how higher-dimensional gravitational blocks work. We will give more details elsewhere \cite{Hosseini:2019HHZ2}.
Notice that, besides recovering known results, our discussion leads to a prediction for the entropy of many rotating higher-dimensional black objects that are still to be found.

\section{Discussion and outlook}
\label{sect:Discussion}

In this paper we provided a general entropy functional that can accommodate all known supersymmetric black holes in AdS$_4\times S^7$ and AdS$_5\times S^5$ and we proposed a generalization to higher dimensions.
Our construction is based on the gluing of gravitational blocks $\cB (X^\Lambda_{(\sigma)} , \omega_{(\sigma)} )$ that is inspired by a field theoretic analogue, the gluing of holomorphic blocks.
As already said many times it would be very interesting to make this analogy more precise, especially because a field theory explanation of some of these results is still missing.

We would also like to stress that there already exist two purely gravitational developments expected to give rise to the same construction. First, Sen's entropy function based on a partially off-shell way of evaluating the supergravity action in the near horizon region of extremal black holes \cite{Sen:2005wa} can in principle be defined for the rotating black holes we consider here. Previous formulations with rotation \cite{Astefanesei:2006dd,Hristov:2018spe} indeed show that Sen's entropy function gets two distinct contributions from the NP and SP of the sphere. However, this construction makes use of real fugacities and is not immediately suited to take into account the constrained Legendre transform of the asymptotically AdS solutions that in general requires complex parameters. Second, the evaluation of the Euclidean on-shell action at the asymptotic boundary of AdS spaces using holographic renormalization is also expected to agree with the entropy functional \cite{Cassani:2019mms}. In particular, it was recently shown \cite{BenettiGenolini:2019jdz} in minimal supergravity that the on-shell action ``localizes'' on isolated fixed points of the supersymmetric Killing vector.%
\footnote{We can also observe some similarity between the contribution from a single fixed point in \cite{BenettiGenolini:2019jdz} and our building block $\cB (X^\Lambda_{(\sigma)} , \omega_{(\sigma)})$ for KN-AdS$_4$ in the minimal supergravity limit.} 
 Since the leading number of degrees of freedom of the black holes is contained within the horizon (known colloquially as the lack of {\it black hole hair}), the asymptotic and the near horizon supergravity actions should agree. We therefore expect a suitable generalization of Sen's entropy function with rotation \cite{Astefanesei:2006dd,Hristov:2018spe} to complex fugacities to agree with a suitable generalization of the ``localization of the action'' of \cite{BenettiGenolini:2019jdz} to non-minimal supergravity, the final answer being given here \eqref{gluing}. 

One can also expect that all these results could follow from an equivariant localization in supergravity along the lines of \cite{Hristov:2019xku,Hristov:2018lod}.
The entropy functional \eqref{gluing} is indeed strongly suggesting an underlying fixed point formula.
Our proposal for a six-dimensional generalization \eqref{conjecture:gluing} is also directly inspired by an equivariant localization computation in field theory.

There are also many other directions for future investigations.

First of all, it would be interesting to consider examples of black objects whose holographically dual SCFT has less supersymmetry.
In particular, there exist static \emph{m}AdS$_4 \times S^6$ black holes in mIIA supergravity \cite{Guarino:2017eag,Guarino:2017pkw} whose effective prepotential reads \cite[(1.2)]{Hosseini:2017fjo}
\be
 \cF ( X^\Lambda ) = - \ii \frac{3^{3/2}}{4} \left( 1 - \frac{\ii}{\sqrt{3}} \right) c^{1/3} ( X^1 X^2 X^3 )^{2/3} \, ,
\ee
where $c$ is the dyonic gauging parameter. The entropy of these black holes has been derived recently in \cite{Azzurli:2017kxo,Hosseini:2017fjo,Benini:2017oxt} via evaluating the topologically twisted of the holographically dual field theory \cite{Guarino:2015jca}.
It would be interesting to find rotating generalization of these black holes and check if our proposal for the attractor mechanism also works in this case.

Second, we notice that our discussion, while focused on AdS$_4$ black holes, has applications also to asymptotically flat black holes.  In particular, the gluing procedure and the associated attractor mechanism can be applied also to
black holes in Mink$_4$. We provide an explicit example in appendix \ref{app:B}.  

We should also note that black holes in gauged supergravity can exist with more exotic horizon topologies,
such as higher genus Riemann surfaces or non-compact hyperbolic space in four dimensions \cite{Caldarelli:1998hg},
and a large number of distinct possibilities when going to higher dimensions. Adding rotation is not possible in every case,
but typically the non-compact horizons do allow for non-vanishing angular momentum.
It would be interesting to extend our findings here to all theses cases as well.

Finally, we can wonder if the gravitational blocks play a bigger role in supergravity. It is  tempting to think that also other supersymmetric observables in gauged supergravity can be evaluated with the help of the building block $\cB (X^\Lambda_{(\sigma)}  , \omega_{(\sigma)} )$, and maybe not just for asymptotically locally AdS backgrounds as appendix \ref{app:B} suggests. Moreover, recalling also that in some cases thermal black holes have been found to follow from a one derivative BPS-like equations \cite{Ceresole:2007wx,Klemm:2012vm,Gnecchi:2012kb}, one might hope to generalize the gravitational blocks to non-supersymmetric cases.

We hope to report more on all these topics in the future.

\section*{Acknowledgements}

We would like to thank Chiung Hwang, Stefanos Katmadas, and Sara Pasquetti for useful discussions.
The work of SMH was supported by World Premier International Research Center Initiative (WPI Initiative), MEXT, Japan.
KH is supported in part by the Bulgarian NSF grants DN08/3 and N28/5.
AZ is partially supported by the INFN, the ERC-STG grant 637844-HBQFTNCER and the MIUR-PRIN contract 2017CC72MK003.

\appendix
\addtocontents{toc}{\protect\setcounter{tocdepth}{1}}

\section[Aspects of 4D \texorpdfstring{$\mathcal{N}=2$}{N=2} supergravity and black hole solutions]{Aspects of 4D ${\cal N}=2$ supergravity and black hole solutions}
\label{app:A}

\subsection{Symplectic vectors and the quartic invariant}
\label{app:A1}
An important symmetry of the equations of motion of supergravity is the electromagnetic duality. As the name suggests, the $n_{\text{V}} + 1$ electric and magnetic gauge field strengths $F^{\Lambda}$ and $G_{\Lambda}$ ($\Lambda = 0, \ldots, n_{\text{V}}$) can be transformed among each other under the symplectic group $\Sp (2 (n_{\text{V}}+1), \mathbb{Z})$, resulting in a rotation of the electromagnetic charges,
\be
	\Gamma = \{ p^\Lambda; q_\Lambda \} \, .
\ee
This needs to be done while simultaneously symplectically rotating a number of other quantities in the theory, such as the gauging parameters 
\be
	G = \{ g^\Lambda ; g_\Lambda \} \, ,
\ee
and the scalars repackaged in special coordinates called symplectic section,
\be
	{\cal V} = e^{{\cal K}/2} \{ X^\Lambda ; F_\Lambda \} \, .
\ee
The ``lower'' part of the symplectic section, $F_\Lambda$, can often be derived from the so-called prepotential ${\cal F} (X)$ by a partial derivative with respect to $X^\Lambda$. The prepotential is a homogeneous function of degree $2$ of the ``upper'' section $X^\Lambda$. In the above formula, ${\cal K}$ is the so-called K\"ahler potential that specifies the metric on the scalar manifold. Note that ${\cal V}$ is uniquely specified by the physical scalars up to a local $\U(1)$ transformation. Conversely, one may always recover the physical scalars by the choice $t^i = X^i / X^0$, $i = 1, \ldots, n_{\text{V}}$.

Inner products of symplectic vectors are denoted by triangle brackets and are naturally defined to be symplectic invariant, \eg\; 
\be
	\Iprod{G}{\Gamma} \equiv g_\Lambda\? p^\Lambda - g^\Lambda\? q_\Lambda \, .
\ee
The section ${\cal V}$ is then subject to the following constraint,
\be
	\Iprod{\bar{\cal V}}{{\cal V}} = \ii \, .
\ee
This fixes the K\"ahler potential ${\cal K}$ and consequently the metric on the scalar manifold from the choice of prepotential ${\cal F} (X)$.
For a complete set of special geometry identities and notations see \cite{Andrianopoli:1996cm}.

A typical example for prepotentials and symplectic rotation is given by the so-called {\it cubic} prepotential
\be\label{eq:cubicprep}
 {\cal F} \big( X^\Lambda \big) =  \frac16 \frac{c_{i j k} X^i X^j X^k}{X^0} \, .
\ee 
with $c_{i j k}$ completely symmetric. Upon symplectic rotation of the vector $\cV$, one can transform the {\it cubic} prepotential into a {\it square root} one,
\be\label{eq:sqrtprep}
 {\cal F} \big( \hat{X}^\Lambda \big) =  2 \ii \sqrt{\hat{X}^0\?\frac16 \hat{c}_{i j k} \hat{X}^i \hat{X}^j \hat{X}^k} \, .
\ee 
where the precise form of the symplectic transformation and the relation between the constant tensors $\hat{c}_{i j k}$ and $c_{i j k}$, as well as between $X^\Lambda$ and $\hat{X}^\Lambda$ can be found in \cite{Gnecchi:2013mta}. Additionally, the scalar manifold resulting from these prepotentials is {\it symmetric} provided the tensors $c, \hat{c}$ satisfy an extra identity, see \eg\ \cite{Hristov:2018spe}. In the main body of this paper we are naturally interested only in string theory embeddings \cite{Cvetic:1999xp} and therefore look at the STU model with non-vanishing $c_{1 2 3} = 1 = \hat{c}_{1 2 3}$ and permutations, such that the scalar manifold is the space $[\SU(1,1)/\U(1)]^3$.

The Lagrangian and supersymmetry variations (and consequently the set of solutions) can be formulated in a manifestly covariant way using the symplectic vectors and their inner products, if one further makes use of the existence of a rank-4 symplectic tensor $t^{M N P Q}$ in the special case of symmetric scalar manifolds \cite{Ferrara:1997uz,Ferrara:2006yb}. The symplectic tensor $t$ is also completely symmetric and it is model-dependent, i.e.\ fixed for a given prepotential ${\cal F}$. In the examples of the cubic and square root prepotentials above, the symplectic tensor $t$ is explicitly given in terms of the tensors $c, \hat{c}$. Upon contraction of this tensor with four different symplectic vectors, \eg\;$\Gamma^{1,2,3,4}$, one defines the so-called quartic invariant form $I_4$ as
\be
	I_4 (\Gamma^1, \Gamma^2, \Gamma^3, \Gamma^4) \equiv t^{M N P Q} \Gamma^1_M \Gamma^2_N \Gamma^3_P \Gamma^4_Q \, .
\ee
where the generalized symplectic indices in the above formula run over both upper and lower $\Lambda$ indices in the previous equations.
One standardly defines the quartic invariant of a single symplectic vector $I_4 (\Gamma)$ with a different symmetry factor,
\be
	I_4 (\Gamma) \equiv \frac{1}{4!}\? t^{M N P Q}\? \Gamma_M \Gamma_N \Gamma_P \Gamma_Q \, ,
\ee
It is also convenient to define as a symplectic vector the first derivative of the quartic invariant,
\be
	I_4' (\Gamma)_M \equiv \Omega_{M N} \frac{\partial I_4 (\Gamma)}{\partial \Gamma_N} \, ,
\ee
where $\Omega_{M N}$ is the inverse of the symplectic form $\Omega^{M N}$. Higher order derivatives and further identities coming from inner products of the quartic invariant with different symplectic vectors can be found in \cite{Bossard:2013oga} and \cite{Halmagyi:2014qza}. 

We note a particularly useful identity following from the properties of the symplectic section ${\cal V}$,
\be
	I_4 ({\rm Re} {\cal V}) = I_4 ({\rm Im} {\cal V}) = \frac{1}{16} \, .
\ee

Finally, let us note that the equations presented in the following two subsections, governing the rotating black holes with and without a twist, can be equally successfully applied to the cases of the general prepotentials \eqref{eq:cubicprep}-\eqref{eq:sqrtprep}. In the main body of this paper we were driven by holography to choose particular string theory embeddings. However, we are confident that the gluing prescription of table \ref{tab:table1} can be applied to arbitrary symmetric models in order to determine the corresponding entropy functionals for different black objects.

\subsection{Rotating black holes with a twist}
\label{app:A2}
Here we are interested in the class of rotating black holes with a twist found in \cite{Hristov:2018spe}. In particular, we focus solely on the near horizon geometry. We summarize the main ingredients and repeat the attractor equations that determines explicitly all the quantities.

Specializing to spherical topology, we can start with the twisting condition that reads
\be
\label{eq:twistingcondition}
	\Iprod{G}{\Gamma} = -1 \, ,
\ee 
where we already made use of the formalism described above and the symplectic vectors for gauging $G$ and electromagnetic charge $\Gamma$.

The metric in the near horizon region is of the form
\be
\rd s^2_4 = -\ex^{2\u} \left(r \? \rd t +  \omega_0 \right)^2 + \ex^{-2\u}\?\left( \frac{\rd r^2}{r^2} + {\rm v}^2 \left( \frac{\rd \theta^2}{\Delta(\theta)} + \Delta(\theta) \sin^2(\theta)\, \rd \phi^2 \right) \! \right) ,
\ee
where
\be
 \ex^{-2\u} = \sqrt{I_4({\cal I}_0)} \, , \quad {\rm v}\?{\cal I}_0 = {\cal H}_0 + {\rm j}\? G \cos (\theta) \,,\quad {\rm v} =  \Iprod{G}{{\cal H}_0} \,  .
\ee
Additionally,
\be
	\Delta (\theta) = 1 - I_4 (G)\? {\rm j}^2 \sin^2 (\theta)\ , \qquad \omega_0 = - \frac{{\rm j}}{{\rm v}}\? \Delta (\theta) \sin^2 (\theta)\? {\rm d} \phi \, ,
\ee
such that the symplectic vector $\cH_0$, together with the extra parameter ${\rm j}$ specify completely the metric.
In the above formulae we already assumed a vanishing NUT charge and the absence of conical singularities near the poles, which further imposes
\be
\label{eq:twistedconstraint}
	\Iprod{\cH_0}{I_4' (G)} = \Iprod{G}{I_4' (\cH_0)} = 0 \, .
\ee
The symplectic sections at the horizon, after a suitable gauge choice, are given by
\be\label{horsec} 
	e^{-{\cal K}/2}\? {\cal V} = \{X^I; F_I \} = - \frac{1}{2 \sqrt{I_4({\cal I}_0)}}\? I_4'({\cal I}_0) + \ii\? {\cal I}_0 \, .
\ee
Ultimately, the solution is uniquely fixed in terms of the conserved electromagnetic charges $\Gamma$ and the angular momentum $\cJ$ from the attractor equations
\be
\label{eq:attr-twisted}
  \Gamma = \frac{1}{4} I^\prime_4\left({\cal H}_0, {\cal H}_0, G \right) + \frac{1}{2}\? {\rm j}^2\? I^\prime_4\left( G \right) ,
\ee
and
\be
 \label{eq:bigJ-twisted}
 {\cal J} = - \frac{{\rm j}}{2} \Big( \Iprod{I_4^\prime(G)}{I_4^\prime(\cH_0)} - \frac12 I_4(\cH_0, \cH_0, G, G) \Iprod{G}{\cH_0} \rule[.1cm]{0pt}{\baselineskip} \Big) \, ,
\ee
that can be used to determine the parameter ${\rm j}$ and the vector $\cH_0$. The allowed conserved charges are however constrained not only by the twisting condition \eqref{eq:twistingcondition}
but also by the constraints \eqref{eq:twistedconstraint} that decrease the parameter space of charges for regular black holes.%
\footnote{Note that in our last example of KN-AdS$_5$ black holes (see section \ref{sec:KNAdS5}), the four-dimensional near horizon solution does indeed have conical singularities near the poles and does not satisfy \eqref{eq:twistedconstraint}.
This is of course physically acceptable, since the five-dimensional uplift is perfectly regular and the apparent singularity in four dimensions is resolved in the uplift.}

It is also useful to define the \emph{real} chemical potential conjugate to the angular momentum $\cJ$ as in \cite{Hristov:2018spe},
\be
\label{eq:twsitedchempotJ}
	w \equiv \frac{{\rm j}}{{\rm v} \sqrt{I_4(\cH_0) - {\rm j}^2}} \, .
\ee

Finally, the quantity of main interest here is the Bekenstein-Hawking entropy, reads
\be
	S_{\text{BH}} = \frac{A}{4 G^{(4)}_{\text{N}}} = \frac{\pi}{G^{(4)}_{\text{N}}} \sqrt{I_4(\cH_0) - {\rm j}^2}\, ,
\ee
which via the attractor equations \eqref{eq:attr-twisted} and \eqref{eq:bigJ-twisted} becomes a function of $\Gamma$ and $\cJ$.

\subsection{Rotating black holes with no twist}
\label{app:A3}

Here instead we focus on the class of black holes \emph{without} a twist, \ie\;the Kerr-Newman-branch recently found in \cite{Hristov:2019mqp}.
Again, we focus purely on the near horizon geometry and the attractor equations that determine fully the solution.

In contrast to the twisting condition in the previous case, \eqref{eq:twistingcondition}, in the present case we have
\be
	\Iprod{G}{\Gamma} = 0 \, .
\ee 

The near horizon metric is given by
\be
\rd s^2_4 = -\ex^{2\u} \left(r  \rd t +  \omega_0 \right)^2 + \ex^{-2\u}\?\left( \ex^{2\sigma_0} \left( \frac{\rd r^2}{R_0^2\? r^2} + \frac{\Xi \rd \theta^2}{\Delta(\theta)} \right) + \frac{R_0^2\? \Delta(\theta)}{\Xi} \sin^2(\theta) \rd \phi^2 \! \right) .
\ee
The various metric functions, as well as the scalars, can eventually be determined by a single symplectic vector $\cC$ in a more convoluted way as compared to before. We have 
\be
 \ex^{-2\u} = \sqrt{I_4({\cal I}_0)} \, , \quad \ex^{2\sigma_0} {\cal I}_0 = {\cal H}_0 \, , \quad \ex^{2\? \sigma_0} =  \frac{\Delta (\theta)}{\Xi} \sin^2 (\theta) + R_0^2 \cos^2 (\theta) \,  ,
\ee
together with
\begin{equation}\label{eq:H0-ansatz-hor}
 \cH_0 = \cC_0 + \cC_1 \cos (\theta) + \cC_2 \cos^2 (\theta) + \cC_3 \cos^3 (\theta) \, .
\end{equation}
We further have
\bea
\cC_0 = & \frac{1}{\Xi} \? \cC\ , \qquad  \cC_1 = \frac{1}{\Xi}\? \left(\Iprod{G}{\cC}\?\cC + \frac{1}{4}\?I_4^\prime(\cC, \cC, G) \right) \, , \qquad && \Xi \equiv 1 - I_4 ( \cC ) I_4( G )  \, , \\
 \cC_2 = & -\frac{1}{2\,\Xi}\? \left( \Iprod{G}{I_4^\prime(\cC)}\? G -\frac{1}{4}\?I_4^\prime(I_4^\prime(\cC), G, G) \right) \, , && \cC_3 = \frac{1}{2\,\Xi}\? I_4(\cC) \? I_4^\prime(G) \, . 
\eea
As before, we can find the symplectic section via the attractor equations
\be\label{XKNAdS4}
	e^{-{\cal K}/2}\? {\cal V} = \{X^I; F_I \} = - \frac{1}{2 \sqrt{I_4({\cal I}_0)}}\? I_4'({\cal I}_0) + \ii\? {\cal I}_0 \, .
\ee
The remaining quantities $\Delta (\theta)$ and $\omega_0$ can also be determined uniquely from the vector $\cC$, see \cite{Hristov:2019mqp}. Without going to further details, we note the constraints
\bea
\label{eq:KNregularity}
	& \Iprod{G}{\cC} = 0 \, , \qquad \Iprod{I_4^\prime(G)}{I_4^\prime(\cC)} = 0 \, , \\
 	& \Xi R_0^2 = 1+I_4(G) I_4(\cC) +\frac{1}{4} I_4 (\cC,\cC,G,G) \, ,
\eea
additionally fixing some of the parameters of the solutions.
The electromagnetic charges and the angular momentum can be obtained via
\be
 \label{attr:eq:KNAdS4}
 \Gamma = \frac{1}{\Xi} \left( \cC + \frac18 I_4' \left( I_4' ( \cC ) , G , G \right) \right) ,
\ee
and%
\footnote{\label{J:rscale:HKT}Here we rescale $\cJ$ by a factor of $-2$ with respect to \cite{Hristov:2019mqp} in order to keep the same normalization in the definition of all conserved charges.}
\be
 \label{JKNAdS4}
 \cJ = -\frac{1}{2 \Xi^2} \left(2\? I_4( \cC ) \left\langle \cC , I_4' ( G ) \right\rangle + \left( 1 + I_4 ( \cC ) I_4 ( G ) \right) \left\langle G , I_4'( \cC ) \right\rangle \right) \, .
\ee
We should note that the explicit form of the attractor equations makes it hard to invert in general the vector $\cC$ in terms of the conserved charges, but a solution can anyway be completely written down.

Finally, the Bekenstein-Hawking entropy is given by
\be\label{SKNAdS4}
 S_{\text{BH}} = \frac{A}{4 G^{(4)}_{\text{N}}}= \frac{\pi}{\Xi G^{(4)}_{\text{N}}} \sqrt{\Xi R_0^2\? I_4 ( \cC ) - \frac{1}{4} \left\langle G , I_4' ( \cC ) \right\rangle^2} \, .
\ee

\section{Rotating black holes in flat space}
\label{app:B}

In a slight digression from the main topic of black holes in AdS, here we discuss the case of asymptotically flat four-dimensional rotating black holes. More precisely, these are the so-called underrotating solutions in \cite{Bossard:2012xsa}, which consist of extremal non-supersymmetric black holes in ungauged supergravity. When seen as solutions of gauged supergravity with vanishing scalar potential, their near horizon geometry however does preserve 2 real supercharges, see \cite{Hristov:2012nu}, and falls inside the general class of rotating horizons with a twist discussed above. Here we show that the attractor mechanism following from the gluing of gravitational blocks holds in full generality for these solutions as well, even if a dual three-dimensional field theory description is lacking and thus the analogy with holomorphic blocks is missing.

As discussed at more length in \cite{Hristov:2018spe}, Minkowski asymptotics in gauged supergravity can be obtained in an arbitrary symmetric cubic model, but here for simplicity we stick to the choice in the main sections, \ie\;the electric STU model
\be
 \cF (X^\Lambda) = \frac{X^1 X^2 X^3}{X^0} \, ,
\ee
and we have the purely electric gauging with a single non-vanishing entry
\be
 G = \{g^{\Lambda}; g_0 , g_{i} \} \, , \qquad g_0 \equiv g \, , \qquad g^\Lambda = g_i = 0 \, .
\ee
We set $g=1$ for further simplicity, but note that here $g$ is not related to the asymptotic length scale (which is of course non-existent in flat space) and therefore one can genuinely consider it as a free parameter, \eg\;coming from a Scherk-Schwarz reduction and further string theory embeddings \cite{Hristov:2014eba}. We can keep a general vector of electromagnetic charges
\be
 \Gamma = \{ p^\Lambda ; q_\Lambda  \} \, ,
\ee
with the twisting condition fixing
\be
	p^0 = -1 \, .
\ee
The full quartic invariant for the electric STU model can again be found in \eqref{eq:I4electricstu}. 

For completeness, since the explicit general formulae use different conventions in the original references, we give the complete near horizon solution here including the auxiliary parameters ${\rm j}, \cH_0$ described in appendix \ref{app:A2}. We first ensure that the NUT charge is vanishing, fixing one of the electric charges in the solution, \eg\;
\be
	q_0 = 2\? p^1 p^2 p^3 + \sum_{i =1}^3 q_i p^i \, .
\ee
We then find the solution
\be
	\cH_0 = \pm \bigg\{1, - p^i; 4 p^1 p^2 p^3 + \sum_{i=1}^3 q_i p^i, q_i + 2\? \frac{p^1 p^2 p^3}{p^i} \bigg\} \, ,
\ee
leading to
\be
	{\rm v} = \pm 1 \, , \qquad {\rm j} = \mp \cJ \, .
\ee
The Bekenstein-Hawking entropy is given by
\be
S_{\text{BH}}^{\text{Mink}} (p^i, q_i, \cJ) = \frac{\pi}{G_{\text{N}}^{(4)}}\? \sqrt{- 4\? \prod_{i=1}^3 \left( q_i + \frac{p^1 p^2 p^3}{p^i} \right) - \cJ^2} \equiv  \frac{\pi}{G^{(4)}_{\text{N}}}\? \Theta \, ,
\ee
and the chemical potential conjugate to the angular momentum becomes
\be
	w = - \frac{\cJ}{\Theta} \, .
\ee
The near horizon values of the sections, evaluated at the North and South poles of the sphere, can be most concisely written as follows:
\bea
 X_{\text{SP}}^\Lambda - X_{\text{NP}}^\Lambda &= - 2 w p^\Lambda \, , \qquad \Lambda = 0, \ldots , 3 \, , \\
 X^0_{\text{SP}} + X^0_{\text{NP}} & = 2 \ii \, , \\
 X_{\text{SP}}^i + X_{\text{NP}}^i & = \frac{2}{\Theta} \bigg( p^i \Big(2 \? p^1 p^2 p^3 + 2 \sum_{j \neq i} q_j p^j - \ii\? \Theta \Big) + 2\? \frac{q_1 q_2 q_3}{q_i} \bigg) , \qquad i = 1, 2, 3 \, .
\eea

\subsection{Attractor mechanism}
The black hole preserves supersymmetry with a topological twist and therefore we need to use the $A$-gluing,
\bea
 X^{\Lambda}_{(1)} & = \chi^\Lambda - \ii\? \omega \? p^\Lambda \, , \qquad && \omega_{(1)} = \omega \, , \\
 X^{\Lambda}_{(2)} & = \chi^\Lambda + \ii\? \omega \? p^\Lambda \, , && \omega_{(2)} = - \omega  \, .
\eea
The constraint on chemical potentials is given by
\be
 \label{const:Mink:chi0}
 \chi^0 =2 \, .
\ee
The entropy functional then reads \eqref{gluing}:
\be
 \label{entropy:functional:Mink}
 \cI_{\text{Mink}} ( p^\Lambda , \chi^\Lambda , \omega ) \equiv  \frac{\pi}{4 G_{\text{N}}^{(4)}}
 \left( \cE_{\text{Mink}} ( p^\Lambda , \chi^\Lambda , \omega ) - 2 \ii \chi^\Lambda q_\Lambda - 2 \omega \cJ \right )
 + \lambda ( \chi^0 - 2) \, ,
\ee
with
\bea
 \cE_{\text{Mink}} ( p^\Lambda , \chi^\Lambda , \omega ) = - \frac{2 \ii}{(4 + \omega^2)}  \bigg(\chi^1 \chi^2 \chi^3 +2 \sum_{i<j<k} \chi^i \chi^j p^k - \omega^2 \Big( \sum_{i < j < k} \chi^ i p^j p^k  + 2\? p^1 p^2 p^3 \Big) \bigg) \, ,
\eea
where for brevity we already used explicitly the constraint for the chemical potential \eqref{const:Mink:chi0} and the twisting condition $p^0 = -1$.

As expected, the proposed attractor mechanism works perfectly. The  values of the sections at the SP $(\theta = 0)$ and the NP $(\theta = \pi)$ are given by
\be
 X^\Lambda_{\text{SP, NP}} =\frac{\ii}{2} \Big( \bar \chi^\Lambda \mp \ii\? \bar{\omega}\? p^\Lambda \Big) \, , \qquad \Lambda = 0 , \ldots, 3 \, ,
\ee
where $\bar \chi$ and $\bar \omega = - 2 w$ are the critical points of the functional $\cI_{\text{Mink}} ( p^i , \chi^\Lambda , \omega )$. Moreover,
\be
 S^{\text{Mink}}_{\text{BH}} ( p^i , q_\Lambda , \cJ) = \cI_{\text{Mink}} ( p^i , \bar \chi^\Lambda , \bar \omega) \, .
\ee

\bibliographystyle{ytphys}

\bibliography{Entropy-Functional-HHZ}

\providecommand{\href}[2]{#2}\begingroup\raggedright\begin{thebibliography}{100}

\bibitem{Benini:2015eyy}
F.~Benini, K.~Hristov, and A.~Zaffaroni, ``{Black hole microstates in AdS$_{4}$
  from supersymmetric localization},''
  \href{http://dx.doi.org/10.1007/JHEP05(2016)054}{{\em JHEP} {\bfseries 05}
  (2016) 054},
\href{http://arxiv.org/abs/1511.04085}{{\ttfamily arXiv:1511.04085 [hep-th]}}.

\bibitem{Cabo-Bizet:2018ehj}
A.~Cabo-Bizet, D.~Cassani, D.~Martelli, and S.~Murthy, ``{Microscopic origin of
  the Bekenstein-Hawking entropy of supersymmetric AdS$_{\bf 5}$ black
  holes},''
\href{http://arxiv.org/abs/1810.11442}{{\ttfamily arXiv:1810.11442 [hep-th]}}.

\bibitem{Choi:2018hmj}
S.~Choi, J.~Kim, S.~Kim, and J.~Nahmgoong, ``{Large AdS black holes from
  QFT},''
\href{http://arxiv.org/abs/1810.12067}{{\ttfamily arXiv:1810.12067 [hep-th]}}.

\bibitem{Benini:2018ywd}
F.~Benini and P.~Milan, ``{Black holes in 4d $\mathcal{N}=4$
  Super-Yang-Mills},''
\href{http://arxiv.org/abs/1812.09613}{{\ttfamily arXiv:1812.09613 [hep-th]}}.

\bibitem{Ferrara:1996dd}
S.~Ferrara and R.~Kallosh, ``{Supersymmetry and attractors},''
  \href{http://dx.doi.org/10.1103/PhysRevD.54.1514}{{\em Phys. Rev.} {\bfseries
  D54} (1996) 1514--1524},
\href{http://arxiv.org/abs/hep-th/9602136}{{\ttfamily arXiv:hep-th/9602136
  [hep-th]}}.

\bibitem{Ferrara:1995ih}
S.~Ferrara, R.~Kallosh, and A.~Strominger, ``{N=2 extremal black holes},''
  \href{http://dx.doi.org/10.1103/PhysRevD.52.R5412}{{\em Phys. Rev.}
  {\bfseries D52} (1995) R5412--R5416},
\href{http://arxiv.org/abs/hep-th/9508072}{{\ttfamily arXiv:hep-th/9508072
  [hep-th]}}.

\bibitem{Ooguri:2004zv}
H.~Ooguri, A.~Strominger, and C.~Vafa, ``{Black hole attractors and the
  topological string},''
  \href{http://dx.doi.org/10.1103/PhysRevD.70.106007}{{\em Phys. Rev.}
  {\bfseries D70} (2004) 106007},
\href{http://arxiv.org/abs/hep-th/0405146}{{\ttfamily arXiv:hep-th/0405146
  [hep-th]}}.

\bibitem{Sen:2005wa}
A.~Sen, ``{Black hole entropy function and the attractor mechanism in higher
  derivative gravity},''
  \href{http://dx.doi.org/10.1088/1126-6708/2005/09/038}{{\em JHEP} {\bfseries
  09} (2005) 038},
\href{http://arxiv.org/abs/hep-th/0506177}{{\ttfamily arXiv:hep-th/0506177
  [hep-th]}}.

\bibitem{Cacciatori:2009iz}
S.~L. Cacciatori and D.~Klemm, ``{Supersymmetric AdS(4) black holes and
  attractors},'' \href{http://dx.doi.org/10.1007/JHEP01(2010)085}{{\em JHEP}
  {\bfseries 01} (2010) 085},
\href{http://arxiv.org/abs/0911.4926}{{\ttfamily arXiv:0911.4926 [hep-th]}}.

\bibitem{DallAgata:2010ejj}
G.~Dall'Agata and A.~Gnecchi, ``{Flow equations and attractors for black holes
  in N = 2 U(1) gauged supergravity},''
  \href{http://dx.doi.org/10.1007/JHEP03(2011)037}{{\em JHEP} {\bfseries 03}
  (2011) 037},
\href{http://arxiv.org/abs/1012.3756}{{\ttfamily arXiv:1012.3756 [hep-th]}}.

\bibitem{Hosseini:2017mds}
S.~M. Hosseini, K.~Hristov, and A.~Zaffaroni, ``{An extremization principle for
  the entropy of rotating BPS black holes in AdS$_{5}$},''
  \href{http://dx.doi.org/10.1007/JHEP07(2017)106}{{\em JHEP} {\bfseries 07}
  (2017) 106},
\href{http://arxiv.org/abs/1705.05383}{{\ttfamily arXiv:1705.05383 [hep-th]}}.

\bibitem{Hosseini:2018dob}
S.~M. Hosseini, K.~Hristov, and A.~Zaffaroni, ``{A note on the entropy of
  rotating BPS AdS$_7\times S^4$ black holes},''
  \href{http://dx.doi.org/10.1007/JHEP05(2018)121}{{\em JHEP} {\bfseries 05}
  (2018) 121},
\href{http://arxiv.org/abs/1803.07568}{{\ttfamily arXiv:1803.07568 [hep-th]}}.

\bibitem{Choi:2018fdc}
S.~Choi, C.~Hwang, S.~Kim, and J.~Nahmgoong, ``{Entropy functions of BPS black
  holes in AdS$_4$ and AdS$_6$},''
\href{http://arxiv.org/abs/1811.02158}{{\ttfamily arXiv:1811.02158 [hep-th]}}.

\bibitem{Cassani:2019mms}
D.~Cassani and L.~Papini, ``{The BPS limit of rotating AdS black hole
  thermodynamics},''
\href{http://arxiv.org/abs/1906.10148}{{\ttfamily arXiv:1906.10148 [hep-th]}}.

\bibitem{Katmadas:2014faa}
S.~Katmadas, ``{Static BPS black holes in U(1) gauged supergravity},''
  \href{http://dx.doi.org/10.1007/JHEP09(2014)027}{{\em JHEP} {\bfseries 09}
  (2014) 027},
\href{http://arxiv.org/abs/1405.4901}{{\ttfamily arXiv:1405.4901 [hep-th]}}.

\bibitem{Halmagyi:2014qza}
N.~Halmagyi, ``{Static BPS black holes in AdS$_{4}$ with general dyonic
  charges},'' \href{http://dx.doi.org/10.1007/JHEP03(2015)032}{{\em JHEP}
  {\bfseries 03} (2015) 032},
\href{http://arxiv.org/abs/1408.2831}{{\ttfamily arXiv:1408.2831 [hep-th]}}.

\bibitem{Hristov:2018spe}
K.~Hristov, S.~Katmadas, and C.~Toldo, ``{Rotating attractors and BPS black
  holes in $AdS_4$},'' \href{http://dx.doi.org/10.1007/JHEP01(2019)199}{{\em
  JHEP} {\bfseries 01} (2019) 199},
\href{http://arxiv.org/abs/1811.00292}{{\ttfamily arXiv:1811.00292 [hep-th]}}.

\bibitem{Benini:2013cda}
F.~Benini and N.~Bobev, ``{Two-dimensional SCFTs from wrapped branes and
  c-extremization},'' \href{http://dx.doi.org/10.1007/JHEP06(2013)005}{{\em
  JHEP} {\bfseries 06} (2013) 005},
\href{http://arxiv.org/abs/1302.4451}{{\ttfamily arXiv:1302.4451 [hep-th]}}.

\bibitem{Hosseini:2019lkt}
S.~M. Hosseini, K.~Hristov, and A.~Zaffaroni, ``{Microstates of rotating
  AdS$_5$ strings},''
\href{http://arxiv.org/abs/1909.08000}{{\ttfamily arXiv:1909.08000 [hep-th]}}.

\bibitem{Cvetic:2005zi}
M.~Cvetic, G.~W. Gibbons, H.~Lu, and C.~N. Pope, ``{Rotating black holes in
  gauged supergravities: Thermodynamics, supersymmetric limits, topological
  solitons and time machines},''
\href{http://arxiv.org/abs/hep-th/0504080}{{\ttfamily arXiv:hep-th/0504080
  [hep-th]}}.

\bibitem{Hristov:2019mqp}
K.~Hristov, S.~Katmadas, and C.~Toldo, ``{Matter-coupled supersymmetric
  Kerr-Newman-AdS$_4$ black holes},''
\href{http://arxiv.org/abs/1907.05192}{{\ttfamily arXiv:1907.05192 [hep-th]}}.

\bibitem{Gutowski:2004ez}
J.~B. Gutowski and H.~S. Reall, ``{Supersymmetric AdS(5) black holes},''
  \href{http://dx.doi.org/10.1088/1126-6708/2004/02/006}{{\em JHEP} {\bfseries
  02} (2004) 006},
\href{http://arxiv.org/abs/hep-th/0401042}{{\ttfamily arXiv:hep-th/0401042
  [hep-th]}}.

\bibitem{Kunduri:2006ek}
H.~K. Kunduri, J.~Lucietti, and H.~S. Reall, ``{Supersymmetric multi-charge
  AdS(5) black holes},''
  \href{http://dx.doi.org/10.1088/1126-6708/2006/04/036}{{\em JHEP} {\bfseries
  04} (2006) 036},
\href{http://arxiv.org/abs/hep-th/0601156}{{\ttfamily arXiv:hep-th/0601156
  [hep-th]}}.

\bibitem{Astefanesei:2006dd}
D.~Astefanesei, K.~Goldstein, R.~P. Jena, A.~Sen, and S.~P. Trivedi,
  ``{Rotating attractors},''
  \href{http://dx.doi.org/10.1088/1126-6708/2006/10/058}{{\em JHEP} {\bfseries
  10} (2006) 058},
\href{http://arxiv.org/abs/hep-th/0606244}{{\ttfamily arXiv:hep-th/0606244
  [hep-th]}}.

\bibitem{Beem:2012mb}
C.~Beem, T.~Dimofte, and S.~Pasquetti, ``{Holomorphic Blocks in Three
  Dimensions},'' \href{http://dx.doi.org/10.1007/JHEP12(2014)177}{{\em JHEP}
  {\bfseries 12} (2014) 177},
\href{http://arxiv.org/abs/1211.1986}{{\ttfamily arXiv:1211.1986 [hep-th]}}.

\bibitem{Belavin:1984vu}
A.~Belavin, A.~Polyakov, and A.~Zamolodchikov, ``Infinite conformal symmetry in
  two-dimensional quantum field theory,''
  \href{http://www.sciencedirect.com/science/article/pii/055032138490052X}{{\em
  Nuclear Physics B} {\bfseries 241} no.~2, (1984) 333 -- 380}.

\bibitem{Nekrasov:2003vi}
N.~A. Nekrasov, ``{Localizing gauge theories},'' in {\em {Mathematical physics.
  Proceedings, 14th International Congress, ICMP 2003, Lisbon, Portugal, July
  28-August 2, 2003}}, pp.~645--654.
\newblock
2003.
\newblock

\bibitem{Pasquetti:2011fj}
S.~Pasquetti, ``{Factorisation of N = 2 Theories on the Squashed 3-Sphere},''
  \href{http://dx.doi.org/10.1007/JHEP04(2012)120}{{\em JHEP} {\bfseries 04}
  (2012) 120},
\href{http://arxiv.org/abs/1111.6905}{{\ttfamily arXiv:1111.6905 [hep-th]}}.

\bibitem{Hwang:2012jh}
C.~Hwang, H.-C. Kim, and J.~Park, ``{Factorization of the 3d superconformal
  index},'' \href{http://dx.doi.org/10.1007/JHEP08(2014)018}{{\em JHEP}
  {\bfseries 08} (2014) 018},
\href{http://arxiv.org/abs/1211.6023}{{\ttfamily arXiv:1211.6023 [hep-th]}}.

\bibitem{Imamura:2013qxa}
Y.~Imamura, H.~Matsuno, and D.~Yokoyama, ``{Factorization of the
  $S^3/\mathbb{Z}_n$ partition function},''
  \href{http://dx.doi.org/10.1103/PhysRevD.89.085003}{{\em Phys. Rev.}
  {\bfseries D89} no.~8, (2014) 085003},
\href{http://arxiv.org/abs/1311.2371}{{\ttfamily arXiv:1311.2371 [hep-th]}}.

\bibitem{Yoshida:2014ssa}
Y.~Yoshida and K.~Sugiyama, ``{Localization of 3d $\mathcal{N}=2$
  Supersymmetric Theories on $S^1 \times D^2$},''
\href{http://arxiv.org/abs/1409.6713}{{\ttfamily arXiv:1409.6713 [hep-th]}}.

\bibitem{Hwang:2015wna}
C.~Hwang and J.~Park, ``{Factorization of the 3d superconformal index with an
  adjoint matter},'' \href{http://dx.doi.org/10.1007/JHEP11(2015)028}{{\em
  JHEP} {\bfseries 11} (2015) 028},
\href{http://arxiv.org/abs/1506.03951}{{\ttfamily arXiv:1506.03951 [hep-th]}}.

\bibitem{Benini:2015noa}
F.~Benini and A.~Zaffaroni, ``{A topologically twisted index for
  three-dimensional supersymmetric theories},''
  \href{http://dx.doi.org/10.1007/JHEP07(2015)127}{{\em JHEP} {\bfseries 07}
  (2015) 127},
\href{http://arxiv.org/abs/1504.03698}{{\ttfamily arXiv:1504.03698 [hep-th]}}.

\bibitem{Nieri:2015yia}
F.~Nieri and S.~Pasquetti, ``{Factorisation and holomorphic blocks in 4d},''
  \href{http://dx.doi.org/10.1007/JHEP11(2015)155}{{\em JHEP} {\bfseries 11}
  (2015) 155},
\href{http://arxiv.org/abs/1507.00261}{{\ttfamily arXiv:1507.00261 [hep-th]}}.

\bibitem{Gukov:2017kmk}
S.~Gukov, D.~Pei, P.~Putrov, and C.~Vafa, ``{BPS spectra and 3-manifold
  invariants},''
\href{http://arxiv.org/abs/1701.06567}{{\ttfamily arXiv:1701.06567 [hep-th]}}.

\bibitem{Closset:2018ghr}
C.~Closset, H.~Kim, and B.~Willett, ``{Seifert fibering operators in 3d
  $\mathcal{N}=2$ theories},''
  \href{http://dx.doi.org/10.1007/JHEP11(2018)004}{{\em JHEP} {\bfseries 11}
  (2018) 004},
\href{http://arxiv.org/abs/1807.02328}{{\ttfamily arXiv:1807.02328 [hep-th]}}.

\bibitem{Hosseini:2016tor}
S.~M. Hosseini and A.~Zaffaroni, ``{Large $N$ matrix models for 3d ${\cal N}=2$
  theories: twisted index, free energy and black holes},''
  \href{http://dx.doi.org/10.1007/JHEP08(2016)064}{{\em JHEP} {\bfseries 08}
  (2016) 064},
\href{http://arxiv.org/abs/1604.03122}{{\ttfamily arXiv:1604.03122 [hep-th]}}.

\bibitem{Hosseini:2016cyf}
S.~M. Hosseini, A.~Nedelin, and A.~Zaffaroni, ``{The Cardy limit of the
  topologically twisted index and black strings in AdS$_{5}$},''
  \href{http://dx.doi.org/10.1007/JHEP04(2017)014}{{\em JHEP} {\bfseries 04}
  (2017) 014},
\href{http://arxiv.org/abs/1611.09374}{{\ttfamily arXiv:1611.09374 [hep-th]}}.

\bibitem{Hosseini:2016ume}
S.~M. Hosseini and N.~Mekareeya, ``{Large $N$ topologically twisted index:
  necklace quivers, dualities, and Sasaki-Einstein spaces},''
  \href{http://dx.doi.org/10.1007/JHEP08(2016)089}{{\em JHEP} {\bfseries 08}
  (2016) 089},
\href{http://arxiv.org/abs/1604.03397}{{\ttfamily arXiv:1604.03397 [hep-th]}}.

\bibitem{Ray:2019lqb}
D.~Jain and A.~Ray, ``{3d $\mathcal{N}=2$ $\widehat{ADE}$ Chern-Simons
  quivers},'' \href{http://dx.doi.org/10.1103/PhysRevD.100.046007}{{\em Phys.
  Rev.} {\bfseries D100} no.~4, (2019) 046007},
\href{http://arxiv.org/abs/1902.10498}{{\ttfamily arXiv:1902.10498 [hep-th]}}.

\bibitem{Wearetookind:2019euv}
D.~Jain, ``{Twisted Indices of more 3d Quivers},''
\href{http://arxiv.org/abs/1908.03035}{{\ttfamily arXiv:1908.03035 [hep-th]}}.

\bibitem{Benini:2016rke}
F.~Benini, K.~Hristov, and A.~Zaffaroni, ``{Exact microstate counting for
  dyonic black holes in AdS4},''
  \href{http://dx.doi.org/10.1016/j.physletb.2017.05.076}{{\em Phys. Lett.}
  {\bfseries B771} (2017) 462--466},
\href{http://arxiv.org/abs/1608.07294}{{\ttfamily arXiv:1608.07294 [hep-th]}}.

\bibitem{Choi:2019zpz}
S.~Choi, C.~Hwang, and S.~Kim, ``{Quantum vortices, M2-branes and black
  holes},''
\href{http://arxiv.org/abs/1908.02470}{{\ttfamily arXiv:1908.02470 [hep-th]}}.

\bibitem{Nian:2019pxj}
J.~Nian and L.~A. Pando~Zayas, ``{Microscopic Entropy of Rotating Electrically
  Charged AdS$_4$ Black Holes from Field Theory Localization},''
\href{http://arxiv.org/abs/1909.07943}{{\ttfamily arXiv:1909.07943 [hep-th]}}.

\bibitem{Chow:2008ip}
D.~D.~K. Chow, ``{Charged rotating black holes in six-dimensional gauged
  supergravity},'' \href{http://dx.doi.org/10.1088/0264-9381/27/6/065004}{{\em
  Class. Quant. Grav.} {\bfseries 27} (2010) 065004},
\href{http://arxiv.org/abs/0808.2728}{{\ttfamily arXiv:0808.2728 [hep-th]}}.

\bibitem{Chow:2007ts}
D.~D.~K. Chow, ``{Equal charge black holes and seven dimensional gauged
  supergravity},'' \href{http://dx.doi.org/10.1088/0264-9381/25/17/175010}{{\em
  Class. Quant. Grav.} {\bfseries 25} (2008) 175010},
\href{http://arxiv.org/abs/0711.1975}{{\ttfamily arXiv:0711.1975 [hep-th]}}.

\bibitem{Hosseini:2018usu}
S.~M. Hosseini, K.~Hristov, A.~Passias, and A.~Zaffaroni, ``{6D attractors and
  black hole microstates},''
  \href{http://dx.doi.org/10.1007/JHEP12(2018)001}{{\em JHEP} {\bfseries 12}
  (2018) 001},
\href{http://arxiv.org/abs/1809.10685}{{\ttfamily arXiv:1809.10685 [hep-th]}}.

\bibitem{Brandhuber:1999np}
A.~Brandhuber and Y.~Oz, ``{The D-4 - D-8 brane system and five-dimensional
  fixed points},'' \href{http://dx.doi.org/10.1016/S0370-2693(99)00763-7}{{\em
  Phys. Lett.} {\bfseries B460} (1999) 307--312},
\href{http://arxiv.org/abs/hep-th/9905148}{{\ttfamily arXiv:hep-th/9905148
  [hep-th]}}.

\bibitem{Hristov:2011ye}
K.~Hristov, C.~Toldo, and S.~Vandoren, ``{On BPS bounds in D=4 N=2 gauged
  supergravity},'' \href{http://dx.doi.org/10.1007/JHEP12(2011)014}{{\em JHEP}
  {\bfseries 12} (2011) 014},
\href{http://arxiv.org/abs/1110.2688}{{\ttfamily arXiv:1110.2688 [hep-th]}}.

\bibitem{Hristov:2013spa}
K.~Hristov, A.~Tomasiello, and A.~Zaffaroni, ``{Supersymmetry on
  Three-dimensional Lorentzian Curved Spaces and Black Hole Holography},''
  \href{http://dx.doi.org/10.1007/JHEP05(2013)057}{{\em JHEP} {\bfseries 05}
  (2013) 057},
\href{http://arxiv.org/abs/1302.5228}{{\ttfamily arXiv:1302.5228 [hep-th]}}.

\bibitem{Cvetic:1999xp}
M.~Cvetic, M.~J. Duff, P.~Hoxha, J.~T. Liu, H.~Lu, J.~X. Lu,
  R.~Martinez-Acosta, C.~N. Pope, H.~Sati, and T.~A. Tran, ``{Embedding AdS
  black holes in ten-dimensions and eleven-dimensions},''
  \href{http://dx.doi.org/10.1016/S0550-3213(99)00419-8}{{\em Nucl. Phys.}
  {\bfseries B558} (1999) 96--126},
\href{http://arxiv.org/abs/hep-th/9903214}{{\ttfamily arXiv:hep-th/9903214
  [hep-th]}}.

\bibitem{Aharony:2008ug}
O.~Aharony, O.~Bergman, D.~L. Jafferis, and J.~Maldacena, ``{N=6 superconformal
  Chern-Simons-matter theories, M2-branes and their gravity duals},''
  \href{http://dx.doi.org/10.1088/1126-6708/2008/10/091}{{\em JHEP} {\bfseries
  10} (2008) 091},
\href{http://arxiv.org/abs/0806.1218}{{\ttfamily arXiv:0806.1218 [hep-th]}}.

\bibitem{Benini:2012cz}
F.~Benini and N.~Bobev, ``{Exact two-dimensional superconformal R-symmetry and
  c-extremization},''
  \href{http://dx.doi.org/10.1103/PhysRevLett.110.061601}{{\em Phys. Rev.
  Lett.} {\bfseries 110} no.~6, (2013) 061601},
\href{http://arxiv.org/abs/1211.4030}{{\ttfamily arXiv:1211.4030 [hep-th]}}.

\bibitem{Gaiotto:2005gf}
D.~Gaiotto, A.~Strominger, and X.~Yin, ``{New connections between 4-D and 5-D
  black holes},'' \href{http://dx.doi.org/10.1088/1126-6708/2006/02/024}{{\em
  JHEP} {\bfseries 02} (2006) 024},
\href{http://arxiv.org/abs/hep-th/0503217}{{\ttfamily arXiv:hep-th/0503217
  [hep-th]}}.

\bibitem{Behrndt:2005he}
K.~Behrndt, G.~Lopes~Cardoso, and S.~Mahapatra, ``{Exploring the relation
  between 4-D and 5-D BPS solutions},''
  \href{http://dx.doi.org/10.1016/j.nuclphysb.2005.10.026}{{\em Nucl. Phys.}
  {\bfseries B732} (2006) 200--223},
\href{http://arxiv.org/abs/hep-th/0506251}{{\ttfamily arXiv:hep-th/0506251
  [hep-th]}}.

\bibitem{Hristov:2014eza}
K.~Hristov, ``{Dimensional reduction of BPS attractors in AdS gauged
  supergravities},'' \href{http://dx.doi.org/10.1007/JHEP12(2014)066}{{\em
  JHEP} {\bfseries 12} (2014) 066},
\href{http://arxiv.org/abs/1409.8504}{{\ttfamily arXiv:1409.8504 [hep-th]}}.

\bibitem{Hristov:2014hza}
K.~Hristov and S.~Katmadas, ``{Wilson lines for AdS$_{5}$ black strings},''
  \href{http://dx.doi.org/10.1007/JHEP02(2015)009}{{\em JHEP} {\bfseries 02}
  (2015) 009},
\href{http://arxiv.org/abs/1411.2432}{{\ttfamily arXiv:1411.2432 [hep-th]}}.

\bibitem{Kapustin:2011jm}
A.~Kapustin and B.~Willett, ``{Generalized Superconformal Index for Three
  Dimensional Field Theories},''
\href{http://arxiv.org/abs/1106.2484}{{\ttfamily arXiv:1106.2484 [hep-th]}}.

\bibitem{Chong:2005hr}
Z.~W. Chong, M.~Cvetic, H.~Lu, and C.~N. Pope, ``{General non-extremal rotating
  black holes in minimal five-dimensional gauged supergravity},''
  \href{http://dx.doi.org/10.1103/PhysRevLett.95.161301}{{\em Phys. Rev. Lett.}
  {\bfseries 95} (2005) 161301},
\href{http://arxiv.org/abs/hep-th/0506029}{{\ttfamily arXiv:hep-th/0506029
  [hep-th]}}.

\bibitem{Kunduri:2007qy}
H.~K. Kunduri and J.~Lucietti, ``{Near-horizon geometries of supersymmetric
  AdS(5) black holes},''
  \href{http://dx.doi.org/10.1088/1126-6708/2007/12/015}{{\em JHEP} {\bfseries
  12} (2007) 015},
\href{http://arxiv.org/abs/0708.3695}{{\ttfamily arXiv:0708.3695 [hep-th]}}.

\bibitem{Kinney:2005ej}
J.~Kinney, J.~M. Maldacena, S.~Minwalla, and S.~Raju, ``{An Index for 4
  dimensional super conformal theories},''
  \href{http://dx.doi.org/10.1007/s00220-007-0258-7}{{\em Commun. Math. Phys.}
  {\bfseries 275} (2007) 209--254},
\href{http://arxiv.org/abs/hep-th/0510251}{{\ttfamily arXiv:hep-th/0510251
  [hep-th]}}.

\bibitem{Romelsberger:2005eg}
C.~Romelsberger, ``{Counting chiral primaries in N = 1, d=4 superconformal
  field theories},''
  \href{http://dx.doi.org/10.1016/j.nuclphysb.2006.03.037}{{\em Nucl. Phys.}
  {\bfseries B747} (2006) 329--353},
\href{http://arxiv.org/abs/hep-th/0510060}{{\ttfamily arXiv:hep-th/0510060
  [hep-th]}}.

\bibitem{Honda:2019cio}
M.~Honda, ``{Quantum Black Hole Entropy from 4d Supersymmetric Cardy
  formula},'' \href{http://dx.doi.org/10.1103/PhysRevD.100.026008}{{\em Phys.
  Rev.} {\bfseries D100} no.~2, (2019) 026008},
\href{http://arxiv.org/abs/1901.08091}{{\ttfamily arXiv:1901.08091 [hep-th]}}.

\bibitem{ArabiArdehali:2019tdm}
A.~Arabi~Ardehali, ``{Cardy-like asymptotics of the 4d $ \mathcal{N}=4 $ index
  and AdS$_{5}$ blackholes},''
  \href{http://dx.doi.org/10.1007/JHEP06(2019)134}{{\em JHEP} {\bfseries 06}
  (2019) 134},
\href{http://arxiv.org/abs/1902.06619}{{\ttfamily arXiv:1902.06619 [hep-th]}}.

\bibitem{Kim:2019yrz}
J.~Kim, S.~Kim, and J.~Song, ``{A 4d $N=1$ Cardy Formula},''
\href{http://arxiv.org/abs/1904.03455}{{\ttfamily arXiv:1904.03455 [hep-th]}}.

\bibitem{Cabo-Bizet:2019osg}
A.~Cabo-Bizet, D.~Cassani, D.~Martelli, and S.~Murthy, ``{The asymptotic growth
  of states of the 4d $ \mathcal{N}=1 $ superconformal index},''
  \href{http://dx.doi.org/10.1007/JHEP08(2019)120}{{\em JHEP} {\bfseries 08}
  (2019) 120},
\href{http://arxiv.org/abs/1904.05865}{{\ttfamily arXiv:1904.05865 [hep-th]}}.

\bibitem{Amariti:2019mgp}
A.~Amariti, I.~Garozzo, and G.~Lo~Monaco, ``{Entropy function from toric
  geometry},''
\href{http://arxiv.org/abs/1904.10009}{{\ttfamily arXiv:1904.10009 [hep-th]}}.

\bibitem{Larsen:2019oll}
F.~Larsen, J.~Nian, and Y.~Zeng, ``{AdS$_5$ Black Hole Entropy near the BPS
  Limit},''
\href{http://arxiv.org/abs/1907.02505}{{\ttfamily arXiv:1907.02505 [hep-th]}}.

\bibitem{Lezcano:2019pae}
A.~G. Lezcano and L.~A. Pando~Zayas, ``{Microstate Counting via Bethe
  Ans\"{a}tze in the 4d ${\cal N}=1$ Superconformal Index},''
\href{http://arxiv.org/abs/1907.12841}{{\ttfamily arXiv:1907.12841 [hep-th]}}.

\bibitem{Lanir:2019abx}
A.~Lanir, A.~Nedelin, and O.~Sela, ``{Black hole entropy function for toric
  theories via Bethe Ansatz},''
\href{http://arxiv.org/abs/1908.01737}{{\ttfamily arXiv:1908.01737 [hep-th]}}.

\bibitem{Andrianopoli:2004im}
L.~Andrianopoli, S.~Ferrara, and M.~A. Lledo, ``{Scherk-Schwarz reduction of D
  = 5 special and quaternionic geometry},''
  \href{http://dx.doi.org/10.1088/0264-9381/21/19/013}{{\em Class. Quant.
  Grav.} {\bfseries 21} (2004) 4677--4696},
\href{http://arxiv.org/abs/hep-th/0405164}{{\ttfamily arXiv:hep-th/0405164
  [hep-th]}}.

\bibitem{Looyestijn:2010pb}
H.~Looyestijn, E.~Plauschinn, and S.~Vandoren, ``{New potentials from
  Scherk-Schwarz reductions},''
  \href{http://dx.doi.org/10.1007/JHEP12(2010)016}{{\em JHEP} {\bfseries 12}
  (2010) 016},
\href{http://arxiv.org/abs/1008.4286}{{\ttfamily arXiv:1008.4286 [hep-th]}}.

\bibitem{Hosseini:2019HHZ2}
S.~M. Hosseini, K.~Hristov, and A.~Zaffaroni, ``{Work in progress},''.

\bibitem{Yoshida:2014qwa}
Y.~Yoshida, ``{Factorization of 4d N=1 superconformal index},''
\href{http://arxiv.org/abs/1403.0891}{{\ttfamily arXiv:1403.0891 [hep-th]}}.

\bibitem{Peelaers:2014ima}
W.~Peelaers, ``{Higgs branch localization of $ \mathcal{N} $ = 1 theories on
  S$^{3}$ x S$^{1}$},'' \href{http://dx.doi.org/10.1007/JHEP08(2014)060}{{\em
  JHEP} {\bfseries 08} (2014) 060},
\href{http://arxiv.org/abs/1403.2711}{{\ttfamily arXiv:1403.2711 [hep-th]}}.

\bibitem{Suh:2018szn}
M.~Suh, ``{Supersymmetric $AdS_6$ black holes from matter coupled $F(4)$ gauged
  supergravity},'' \href{http://dx.doi.org/10.1007/JHEP02(2019)108}{{\em JHEP}
  {\bfseries 02} (2019) 108},
\href{http://arxiv.org/abs/1810.00675}{{\ttfamily arXiv:1810.00675 [hep-th]}}.

\bibitem{Hosseini:2018uzp}
S.~M. Hosseini, I.~Yaakov, and A.~Zaffaroni, ``{Topologically twisted indices
  in five dimensions and holography},''
  \href{http://dx.doi.org/10.1007/JHEP11(2018)119}{{\em JHEP} {\bfseries 11}
  (2018) 119},
\href{http://arxiv.org/abs/1808.06626}{{\ttfamily arXiv:1808.06626 [hep-th]}}.

\bibitem{Crichigno:2018adf}
P.~M. Crichigno, D.~Jain, and B.~Willett, ``{5d Partition Functions with A
  Twist},'' \href{http://dx.doi.org/10.1007/JHEP11(2018)058}{{\em JHEP}
  {\bfseries 11} (2018) 058},
\href{http://arxiv.org/abs/1808.06744}{{\ttfamily arXiv:1808.06744 [hep-th]}}.

\bibitem{Bawane:2014uka}
A.~Bawane, G.~Bonelli, M.~Ronzani, and A.~Tanzini, ``{$\mathcal{N}=2$
  supersymmetric gauge theories on $S^2\times S^2$ and Liouville Gravity},''
  \href{http://dx.doi.org/10.1007/JHEP07(2015)054}{{\em JHEP} {\bfseries 07}
  (2015) 054},
\href{http://arxiv.org/abs/1411.2762}{{\ttfamily arXiv:1411.2762 [hep-th]}}.

\bibitem{Bershtein:2015xfa}
M.~Bershtein, G.~Bonelli, M.~Ronzani, and A.~Tanzini, ``{Exact results for $
  \mathcal{N} $ = 2 supersymmetric gauge theories on compact toric manifolds
  and equivariant Donaldson invariants},''
  \href{http://dx.doi.org/10.1007/JHEP07(2016)023}{{\em JHEP} {\bfseries 07}
  (2016) 023},
\href{http://arxiv.org/abs/1509.00267}{{\ttfamily arXiv:1509.00267 [hep-th]}}.

\bibitem{Seiberg:1996bd}
N.~Seiberg, ``{Five-dimensional SUSY field theories, nontrivial fixed points
  and string dynamics},''
  \href{http://dx.doi.org/10.1016/S0370-2693(96)01215-4}{{\em Phys. Lett.}
  {\bfseries B388} (1996) 753--760},
\href{http://arxiv.org/abs/hep-th/9608111}{{\ttfamily arXiv:hep-th/9608111
  [hep-th]}}.

\bibitem{Jafferis:2012iv}
D.~L. Jafferis and S.~S. Pufu, ``{Exact results for five-dimensional
  superconformal field theories with gravity duals},''
  \href{http://dx.doi.org/10.1007/JHEP05(2014)032}{{\em JHEP} {\bfseries 05}
  (2014) 032},
\href{http://arxiv.org/abs/1207.4359}{{\ttfamily arXiv:1207.4359 [hep-th]}}.

\bibitem{DAuria:2000xty}
R.~D'Auria, S.~Ferrara, and S.~Vaula, ``{F(4) supergravity and 5-D
  superconformal field theories},''
  \href{http://dx.doi.org/10.1088/0264-9381/18/16/308}{{\em Class. Quant.
  Grav.} {\bfseries 18} (2001) 3181--3196},
\href{http://arxiv.org/abs/hep-th/0008209}{{\ttfamily arXiv:hep-th/0008209
  [hep-th]}}.

\bibitem{Andrianopoli:2001rs}
L.~Andrianopoli, R.~D'Auria, and S.~Vaula, ``{Matter coupled F(4) gauged
  supergravity Lagrangian},''
  \href{http://dx.doi.org/10.1088/1126-6708/2001/05/065}{{\em JHEP} {\bfseries
  05} (2001) 065},
\href{http://arxiv.org/abs/hep-th/0104155}{{\ttfamily arXiv:hep-th/0104155
  [hep-th]}}.

\bibitem{BenettiGenolini:2019jdz}
P.~Benetti~Genolini, J.~M. P\'erez Ipi\~{n}a, and J.~Sparks, ``{Localization of
  the action in AdS/CFT},''
\href{http://arxiv.org/abs/1906.11249}{{\ttfamily arXiv:1906.11249 [hep-th]}}.

\bibitem{Hristov:2019xku}
K.~Hristov, I.~Lodato, and V.~Reys, ``{One-loop determinants for black holes in
  4d gauged supergravity},''
\href{http://arxiv.org/abs/1908.05696}{{\ttfamily arXiv:1908.05696 [hep-th]}}.

\bibitem{Hristov:2018lod}
K.~Hristov, I.~Lodato, and V.~Reys, ``{On the quantum entropy function in 4d
  gauged supergravity},'' \href{http://dx.doi.org/10.1007/JHEP07(2018)072}{{\em
  JHEP} {\bfseries 07} (2018) 072},
\href{http://arxiv.org/abs/1803.05920}{{\ttfamily arXiv:1803.05920 [hep-th]}}.

\bibitem{Guarino:2017eag}
A.~Guarino and J.~Tarrio, ``{BPS black holes from massive IIA on S$^{6}$},''
  \href{http://dx.doi.org/10.1007/JHEP09(2017)141}{{\em JHEP} {\bfseries 09}
  (2017) 141},
\href{http://arxiv.org/abs/1703.10833}{{\ttfamily arXiv:1703.10833 [hep-th]}}.

\bibitem{Guarino:2017pkw}
A.~Guarino, ``{BPS black hole horizons from massive IIA},''
  \href{http://dx.doi.org/10.1007/JHEP08(2017)100}{{\em JHEP} {\bfseries 08}
  (2017) 100},
\href{http://arxiv.org/abs/1706.01823}{{\ttfamily arXiv:1706.01823 [hep-th]}}.

\bibitem{Hosseini:2017fjo}
S.~M. Hosseini, K.~Hristov, and A.~Passias, ``{Holographic microstate counting
  for AdS$_{4}$ black holes in massive IIA supergravity},''
  \href{http://dx.doi.org/10.1007/JHEP10(2017)190}{{\em JHEP} {\bfseries 10}
  (2017) 190},
\href{http://arxiv.org/abs/1707.06884}{{\ttfamily arXiv:1707.06884 [hep-th]}}.

\bibitem{Azzurli:2017kxo}
F.~Azzurli, N.~Bobev, P.~M. Crichigno, V.~S. Min, and A.~Zaffaroni, ``{A
  universal counting of black hole microstates in AdS$_{4}$},''
  \href{http://dx.doi.org/10.1007/JHEP02(2018)054}{{\em JHEP} {\bfseries 02}
  (2018) 054},
\href{http://arxiv.org/abs/1707.04257}{{\ttfamily arXiv:1707.04257 [hep-th]}}.

\bibitem{Benini:2017oxt}
F.~Benini, H.~Khachatryan, and P.~Milan, ``{Black hole entropy in massive Type
  IIA},'' \href{http://dx.doi.org/10.1088/1361-6382/aa9f5b}{{\em Class. Quant.
  Grav.} {\bfseries 35} no.~3, (2018) 035004},
\href{http://arxiv.org/abs/1707.06886}{{\ttfamily arXiv:1707.06886 [hep-th]}}.

\bibitem{Guarino:2015jca}
A.~Guarino, D.~L. Jafferis, and O.~Varela, ``{String Theory Origin of Dyonic
  N=8 Supergravity and Its Chern-Simons Duals},''
  \href{http://dx.doi.org/10.1103/PhysRevLett.115.091601}{{\em Phys. Rev.
  Lett.} {\bfseries 115} no.~9, (2015) 091601},
\href{http://arxiv.org/abs/1504.08009}{{\ttfamily arXiv:1504.08009 [hep-th]}}.

\bibitem{Caldarelli:1998hg}
M.~M. Caldarelli and D.~Klemm, ``{Supersymmetry of Anti-de Sitter black
  holes},'' \href{http://dx.doi.org/10.1016/S0550-3213(98)00846-3}{{\em Nucl.
  Phys.} {\bfseries B545} (1999) 434--460},
\href{http://arxiv.org/abs/hep-th/9808097}{{\ttfamily arXiv:hep-th/9808097
  [hep-th]}}.

\bibitem{Ceresole:2007wx}
A.~Ceresole and G.~Dall'Agata, ``{Flow Equations for Non-BPS Extremal Black
  Holes},'' \href{http://dx.doi.org/10.1088/1126-6708/2007/03/110}{{\em JHEP}
  {\bfseries 03} (2007) 110},
\href{http://arxiv.org/abs/hep-th/0702088}{{\ttfamily arXiv:hep-th/0702088
  [hep-th]}}.

\bibitem{Klemm:2012vm}
D.~Klemm and O.~Vaughan, ``{Nonextremal black holes in gauged supergravity and
  the real formulation of special geometry II},''
  \href{http://dx.doi.org/10.1088/0264-9381/30/6/065003}{{\em Class. Quant.
  Grav.} {\bfseries 30} (2013) 065003},
\href{http://arxiv.org/abs/1211.1618}{{\ttfamily arXiv:1211.1618 [hep-th]}}.

\bibitem{Gnecchi:2012kb}
A.~Gnecchi and C.~Toldo, ``{On the non-BPS first order flow in N=2 U(1)-gauged
  Supergravity},'' \href{http://dx.doi.org/10.1007/JHEP03(2013)088}{{\em JHEP}
  {\bfseries 03} (2013) 088},
\href{http://arxiv.org/abs/1211.1966}{{\ttfamily arXiv:1211.1966 [hep-th]}}.

\bibitem{Andrianopoli:1996cm}
L.~Andrianopoli, M.~Bertolini, A.~Ceresole, R.~D'Auria, S.~Ferrara, P.~Fre, and
  T.~Magri, ``{N=2 supergravity and N=2 superYang-Mills theory on general
  scalar manifolds: Symplectic covariance, gaugings and the momentum map},''
  \href{http://dx.doi.org/10.1016/S0393-0440(97)00002-8}{{\em J. Geom. Phys.}
  {\bfseries 23} (1997) 111--189},
\href{http://arxiv.org/abs/hep-th/9605032}{{\ttfamily arXiv:hep-th/9605032
  [hep-th]}}.

\bibitem{Gnecchi:2013mta}
A.~Gnecchi and N.~Halmagyi, ``{Supersymmetric black holes in $AdS_4$ from very
  special geometry},'' \href{http://dx.doi.org/10.1007/JHEP04(2014)173}{{\em
  JHEP} {\bfseries 04} (2014) 173},
\href{http://arxiv.org/abs/1312.2766}{{\ttfamily arXiv:1312.2766 [hep-th]}}.

\bibitem{Ferrara:1997uz}
S.~Ferrara and M.~Gunaydin, ``{Orbits of exceptional groups, duality and BPS
  states in string theory},''
  \href{http://dx.doi.org/10.1142/S0217751X98000913}{{\em Int. J. Mod. Phys.}
  {\bfseries A13} (1998) 2075--2088},
\href{http://arxiv.org/abs/hep-th/9708025}{{\ttfamily arXiv:hep-th/9708025
  [hep-th]}}.

\bibitem{Ferrara:2006yb}
S.~Ferrara, E.~G. Gimon, and R.~Kallosh, ``{Magic supergravities, N= 8 and
  black hole composites},''
  \href{http://dx.doi.org/10.1103/PhysRevD.74.125018}{{\em Phys. Rev.}
  {\bfseries D74} (2006) 125018},
\href{http://arxiv.org/abs/hep-th/0606211}{{\ttfamily arXiv:hep-th/0606211
  [hep-th]}}.

\bibitem{Bossard:2013oga}
G.~Bossard and S.~Katmadas, ``{Duality covariant multi-centre black hole
  systems},'' \href{http://dx.doi.org/10.1007/JHEP08(2013)007}{{\em JHEP}
  {\bfseries 08} (2013) 007},
\href{http://arxiv.org/abs/1304.6582}{{\ttfamily arXiv:1304.6582 [hep-th]}}.

\bibitem{Bossard:2012xsa}
G.~Bossard and S.~Katmadas, ``{Duality covariant non-BPS first order
  systems},'' \href{http://dx.doi.org/10.1007/JHEP09(2012)100}{{\em JHEP}
  {\bfseries 09} (2012) 100},
\href{http://arxiv.org/abs/1205.5461}{{\ttfamily arXiv:1205.5461 [hep-th]}}.

\bibitem{Hristov:2012nu}
K.~Hristov, S.~Katmadas, and V.~Pozzoli, ``{Ungauging black holes and hidden
  supercharges},'' \href{http://dx.doi.org/10.1007/JHEP01(2013)110}{{\em JHEP}
  {\bfseries 01} (2013) 110},
\href{http://arxiv.org/abs/1211.0035}{{\ttfamily arXiv:1211.0035 [hep-th]}}.

\bibitem{Hristov:2014eba}
K.~Hristov and A.~Rota, ``{6d-5d-4d reduction of BPS attractors in flat gauged
  supergravities},''
  \href{http://dx.doi.org/10.1016/j.nuclphysb.2015.05.023}{{\em Nucl. Phys.}
  {\bfseries B897} (2015) 213--228},
\href{http://arxiv.org/abs/1410.5386}{{\ttfamily arXiv:1410.5386 [hep-th]}}.

\end{thebibliography}\endgroup

\end{document}